\documentclass[psfig]{mn2e}
\usepackage{graphicx}
\title[Spectral $k$-corrections and the colour-magnitude relation]
 {Spectral-based $k$-corrections and implications for the colour-magnitude relation of E/S0s and its evolution} 
\author[N. Roche, M. Bernardi and J. Hyde] 
 {Nathan Roche, Mariangela Bernardi and Joseph Hyde 
 \thanks{nathanroche@mac.com; bernardi@sas.upenn.edu; jhyde@sas.upenn.edu}\\
  University of Pennsylvania, 209 South 33rd Street, 
  Physics and Astronomy, Philadelphia, PA 19104, USA}

\begin{document}

\date{}

\pagerange{\pageref{firstpage}--\pageref{lastpage}} \pubyear{2009}

\maketitle

\label{firstpage}

\begin{abstract} 
We select a sample of 70378 E/S0 (early-type) galaxies at $0<z<0.36$ from the Sloan Digital Sky Survey, excluding disk and star-forming galaxies. We estimate $g$ and $r$ magnitudes in the observer- and rest-frames directly from the SDSS DR6 spectra; this provides an object-by-object estimate of the $k$-correction.  Observer-frame colours can also be estimated from the imaging photometry.  However, in this case, rest-frame colours, and hence $k$-corrections, must be inferred from fitting to data in other band-passes as well as to stellar population synthesis models.  There are small (a few 0.01 magnitudes) discrepancies between the spectra and imaging photometry, particularly for galaxies with with low signal/noise ratios in the spectra and more negative {\tt eclass} spectral classifications. We correct for these, and then use the $k$-corrections from the spectra to study the evolution of the rest-frame colour-magnitude (CMR) and colour-$\sigma$ (C$\sigma$R) relations.  

Both the CMR and C$\sigma$R relations evolve blueward with increasing redshift, approximately in agreement with passive evolution models with age $\sim 12$ Gyr. The rate of evolution is sensitive to the $k$-corrections.  Using $k$-corrections from the CWW template spectrum, rather than the observed SDSS spectra, gives a CMR with much more evolution; $k$-corrections from Blanton \& Roweis, which are based on fitting Bruzual \& Charlot models with solar abundance ratios to the $ugriz$ colours, give less/no evolution.  However, the slope and zero-point of CMR relation depends on whether colours were defined in fixed physical or angular apertures, a consequence of the fact that the centers of these objects tend to be redder: the relation is steeper for fixed angular apertures.  These colour gradients must be accounted for when estimating CMR evolution with fixed angular apertures.  On the other hand, although the zero-point of the C$\sigma$R depends on the aperture in which the colour was defined, the slope does not, suggesting that colour gradients are correlated with residuals from the $\sigma-M_r$ relation.  Since these residuals are age indicators, our findings suggest that colour gradients depend on the age of the stellar population.

\end{abstract}


\section{Introduction} 
It has long been known (e.g. Visvanathan and Sandage 1977) that the more luminous early-type (elliptical and S0) galaxies tend to have redder colours. Expressed in magnitudes, the colour-magnitude relation (CMR) is approximately linear. The form of the CMR is most readily observed in a rich galaxy cluster, which provides a very wide luminosity range of E/S0s at a single redshift.

 The CMR provides significant information about galaxy evolution. Firstly, from the tightness  (scatter $<0.05$ mag) of the CMR for the E/S0s in the Coma and Virgo clusters, Bower, Lucey and Ellis (1992) concluded that these galaxies had formed at $z>2$.  Secondly, the slope of the CMR appeared to derive primarily from a correlation of metallicity with galaxy mass (e.g. Kodama and Aromoto 1997; Vazdekis et al. 2001). Higher metallicity results in redder colours for a given age. The observed trend would reflect the greater ability of more massive E/S0s to retain metal-enriched gas during their formative starbursts. An increase in stellar age with galaxy mass may play a secondary role.

A `red sequence' of E/S0 galaxies is seen in rich clusters out to at least $z=1.27$, with little or no change in the CMR slope (Blakeslee et al. 2003; Mei et al. 2006, 2009). The red sequence  appears to have first formed at $z\sim 2$ (Labb\'e et al. 2007). For a passively aging stellar population there would be a blueward shift of the CMR with increasing lookback time (redshift); Holden et al. (2004) and Taylor et al. (2008) find some evidence of this colour evolution. 

In this paper we aim to obtain an accurate measurement of the evolution of the CMR to $z=0.36$. However, to determine the CMR over a wide redshift range, but in a fixed rest-frame colour (e.g. $g-r$), we need to apply accurate $k$-corrections to convert the observed photometric magnitudes back to the galaxy rest-frames. Galaxy $k$-corrections are commonly derived from modelled spectra, but in reality each galaxy has its own individual $k$-correction, and the uncertainties in modelled $k$-corrections (even a few 0.01 mag) may well be comparable to the small evolution that we aim to study in the CMR.  Thus, it would be much preferable to derive either $k$-corrections, or rest-frame colours,  directly from the galaxy spectra, but for good statistics on the CMR a database of many thousand high quality, flux calibrated spectra are required.

The Sloan Digital Sky Survey (SDSS, Stoughton et al. 2002) had as one of 
its stated aims to revolutionize the study of galaxies at moderate 
redshifts, by obtaining hundreds of thousands of spectra together with 
five-band ($ugriz$) photometry. Bernardi et al. (2003a) carried out an 
initial study of the early-type galaxies in the SDSS, selecting 9000 E/S0s 
(from the $\sim 65000$ galaxy spectra available at that time) using a 
combination of morphological and spectral criteria. 
Rest-frame colours and magnitudes were estimated, using $k$-corrections 
based on
 (i) an evolving Bruzual \& Charlot (2003, hereafter BC03) 
 model with a single burst of star-formation 9 Gyr ago, with solar 
 abundances and a Kroupa (2001) IMF,
 (ii) an observed elliptical template spectrum from 
 Coleman, Wu and Weedman (1980, hereafter CWW). 

For these E/S0 galaxies Bernardi et al. (2003b) estimated evolution 
rates of $\Delta(M_g)=-1.15z$, $\Delta(M_r)=-0.85z$ and 
$\Delta(M_i)=-0.75z$. Bernardi et al. (2003c) derived a CMR  
using a k-correction from a composite of the above two models, 
$K_{CWW}^{evol}(z)=K^{no-ev}_{CWW}(z)+K^{evol}_{BC}-K^{no-ev}_{BC}$. 
The rest-frame colour $(g-r)_{rf}$ correlated with luminosity and 
velocity dispersion as $(-0.025\pm 0.003)M_r$ and 
$\sigma^{0.26\pm 0.02}$. Metal abundances in the E/S0s correlated 
with velocity dispersion e.g. as ${\rm Mg_2}\propto \sigma^{0.20\pm 0.02}$ 
and $\langle \rm Fe \rangle \propto \sigma^{0.11\pm 0.02}$, and 
the mean stellar age was estimated as 9 Gyr.
  
Bernardi et al. (2005) re-examined the CMR, using 39320 early-type galaxies from the SDSS, and showed that it was the result of more fundamental correlations between colour and velocity dispersion $\sigma$, and between $\sigma$ and luminosity.  %The argument for this was that, at a constant $M_r^*$, redder colour was found to correlate with higher $\sigma$, whereas at constant $\sigma$ the colour did not correlate with $M_r^*$.

Gallazzi et al (2006), using a similar sample of SDSS E/S0s, found a steep correlation of both metallicity and Mg/Fe ratio with stellar mass, as traced by either luminosity or velocity dispersion (e.g. ${\rm Mg_2}\propto \sigma^{0.25\pm0.02}$). They explained the CMR as a sequence in stellar mass, with some increase in mean age. Their CMR exhibited only a minimal redward shift ($0.006\pm0.003$ mag in $g-r$) between less and more dense clustering environments.  Jimenez et al. (2007) concluded from the small scatter in the CMR that the E/S0s formed most of their stars at $z>2$, and from the super-solar Mg/Fe abundance ratios (`$\alpha$-enhancement') of the more massive E/S0s that their star-formation took place in a short burst ($\sim 1$ Gyr). 

Continued and ongoing SDSS observations have greatly enlarged the database of spectra, enabling us to analyse the CMR of a larger sample of E/S0s than previously. In Section 2 we describe the data and sample selection, and in Section 3, the derivation of the $k$-corrections from the spectra. In Section 4 we present our CMR and compare it with CMRs obtained using other apertures and $k$-correction prescriptions.  We present a similar analysis of the C$\sigma$R in Section 5. 
%In Section 6  we examine the residuals of the E/S0 colours relative to these relations, in Section 7 compare the CMR of the BCGs.
We summarize and discuss our results in Section 6. 

SDSS magnitudes are given in the AB system where
 $m_{AB}=-48.60-2.5$ log $F_{\nu}$ and $m_{AB}=0$ is 3631 Jy.
We assume throughout a spatially flat cosmology with
 $H_0=70$ km $\rm s^{-1}Mpc^{-1}$, 
 $\Omega_{M}=0.27$ and $\Omega_{\Lambda}=0.73$, giving 
the age of the Universe as 13.88 Gyr. 

\section{SDSS Data}
\subsection{Selecting the E/S0 sample}
 Our sample is selected from Data Release 4 of the Sloan Digital 
Sky Survey (Adelman-McCarthy et al. 2006), which covered 4783 $\rm deg^2$ 
of sky with spectra for a total of 673280 sources. However, our 
analysis uses more recent versions of the spectra from SDSS Data 
Release 6 (Adelman-McCarthy et al. 2008), calibrated at Princeton 
University (spectro.princeton.edu). These reductions include 
measurements of the line-of-signt velocity dispersion ($\sigma$) 
through the 3 arcsec diameter aperture.

 This dataset contained 367471 galaxies of all types (but excluding 
stars, QSOs and unclassifiable sources) in the range $14.5<r<17.77$ 
(Petrosian AB magnitudes) with  flux/wavelength calibrated optical 
spectra (approx. 3800--9200$\rm \AA$ with resolution 1800-2100) and 
$ugriz$ photometry.
The SDSS image processing software provides several sets of $ugriz$ 
magnitudes for each object. The photometric systems of interest here 
are:\\
(i) {\tt Model} magnitudes: the total magnitude corresponding to the best-fit profile, which is for these galaxies a de Vaucouleurs (1948) profile, with an effective radius ($r_{deV}$) fitted in the $r$ band. The radius $r_{deV}$ is kept fixed for the magnitude measurement in the other passbands, so as to give accurate colours even for galaxies with a strong colour gradient.\\
(ii) {\tt Fiber} magnitudes: measured in fixed 3 arcsec apertures on the imaging data after smoothing to a uniform 2 arcsec resolution -- intended to correspond to the spectroscopic fiber aperture.\\
(iii) Petrosian magnitudes. The Petrosian (1976) radius $r_{Pet}$ is the angular radius where the mean surface brightness averaged over the aperture $r<r_{Pet}$ is 5 times higher than the mean surface brightness on the annulus $r=r_{Pet}$. The Petrosian flux/magnitude is the total flux observed within twice this radius.\\
We apply strict selection criteria in order to extract a clean sample of E/S0s with good spectra, and exclude spirals, galaxies with disk components, and star-forming galaxies. The most important selection criteria are that:\\
(i) $ \rm frac_{deV}(r)$: the fraction of total flux contributed by the de Vaucouleurs component  when the galaxy profile is fitted by the sum of de Vaucouleurs and exponential profiles. We include only non-disk galaxies with $\rm frac_{deV}$ of unity for both the $g$ and the $r$ band. \\
(ii) {\tt Eclass}, a 1D classification of spectral type, obtained from a principal component analysis (Yip et al. 2004) -- lower (negative) values corresponding to absorption-line galaxies with old stellar populations, positive values indicating star-forming galaxies. We include only galaxies with $\rm {\tt eclass}<0$.\\
In addition we require that:\\
(iii) Concentration, defined as the ratio of the Petrosian radius $r_{Pet, 90}$ (the angular radius containing $90\%$ of the Petrosian flux) to the smaller radius $r_{Pet,50}$ (containing $50\%$), measured in the $i$-band, must be $>2.4$ to reduce possible contamination of later-type galaxies (this selection excludes only a few objects).\\
(iv) The $r$ band {\tt model} magnitude, corrected for Galactic reddening, must be $14.5<r<17.5$ (this excludes some of the faintest objects).\\
(v) The velocity dispersion ($\sigma$) as measured from the SDSS spectrum must be at least 60 km $\rm s^{-1}$, but no more than 450 km $\rm s^{-1}$ (the upper limit to exclude superimposed galaxies, as no single galaxy in the SDSS appears to have $\sigma>450$ km $s^{-1}$; e.g Bernardi et al. 2008; Salviander et al. 2008)\\
 (vi) Redshift must be $0.005\leq z \leq 0.36$. The reason for the upper limit is that, at $z>0.36$, the rest-frame $r$-band begins to shift significantly outside of the 3800--$\rm 9200\AA$ range of the SDSS spectra, so we cannot obtain an accurate $r$-band k-correction. \\
The selected galaxies, satisfying all the above, number 70378.

\subsection{Corrections applied to SDSS parameters}
We apply a number of, generally very small, corrections to the $\sigma$, radii and model magnitudes given by the SDSS catalogs.

The  $\sigma$ measured for early-type galaxies decreases (very slowly) with distance from the centre, and hence the SDSS $\sigma$ will have a dependency on the ratio of the spectrograph aperture to the galaxy's effective  radius.
J{\/o}rgensen et al. (1995) estimated that $\sigma\propto (r_{ap}/r_{deV})^{-0.04}$, where $r_{deV}$ is the de Vaucouleurs model radius in the $r$-band. As in Bernardi et al. (2003a), we correct all the $\sigma$ measurements from the SDSS spectra to apertures of $r_{deV}/8$. As $r_{ap}=1.5$ arcsec for the SDSS, and $r_{deV}$ is in arcsec,  this gives
 $\sigma_{corr}= \sigma_{SDSS}(r_{deV}/12.0)^{-0.04}$.

We correct the de Vaucouleurs model-fit radii, which are semi-major axes, to effective radii ($r_{eff}$) by multiplying by the square root of the  axis ratio, $r_{eff}=r_{deV}\sqrt{b/a}$.

Furthermore, there is evidence of systematic errors in the sky subtraction in the SDSS reductions (e.g. Bernardi et al. 2007; Lauer et al. 2007; Hyde \& Bernardi 2009). These caused the size and total flux of large, extended objects to be underestimated. By comparing SDSS catalogs and more accurate reductions, Hyde \& Bernardi (2009) estimated corrections for the SDSS effective radii. These were positive,  a steep function of $r_{eff}$, and zero for $r_{eff}\leq2.0$ arcsec. Corrections were also estimated for the de Vaucouleurs model fit magnitudes, these being also a function of $r_{eff}$, but zero for $r_{eff}\leq 1.5$ arcsec. 

We correct all our $r_{eff}$ radii as in Hyde \& Bernardi equation 4. The {\tt model} magnitudes, for these galaxies with $\rm frac_{deV}=1.0$, are the same as de Vaucouleurs model-fit magnitudes, except that the radius of the $r$-band fit is used for all passbands. Hence we correct our {\tt model} magnitudes in all ($ugriz$) passbands as in equation 3 of 
Hyde \& Bernardi (2009) but always using the $r$-band $r_{eff}$.
  This correction will therefore cause no change in the model-magnitude  colours. For the {\tt fiber} magnitudes, no sky-subtraction correction is applied, as these small aperture magnitudes correspond approximately to a model fit with $r_{eff}=1.5$ arcsec, where the this correction falls to zero.

\section{Deriving $k$-corrections from spectra} 
\subsection{Method}
We calculate $g$ and $r$-band $k$-corrections directly from the flux-calibrated spectra, corrected for atmospheric and Galactic reddening.  To determine the k-correction for a galaxy at redshift $z$, we first integrate the galaxy spectrum $F(\lambda)$ over the filter response function. In the $g$-band, $g(\lambda)$, the apparent magnitude $g_{spec}$, in the AB system, is (Hogg 2002; Bruzual \& Charlot 2003)
\begin{equation}
 g_{spec} = -2.5~{\rm log10}
  \frac{\int F(\lambda)g(\lambda)\lambda^2\, d\ln\lambda}
       {\int C(\lambda)g(\lambda)\lambda^2\, d\ln\lambda}
\end{equation}
where
 $C(\lambda)\equiv 3631\, (c/\lambda^2)\times 10^{-23}$~erg~s$^{-1}$~cm$^{-2}$
is the spectrum of a source with $F_{\nu}=3631.0$ Jy at all frequencies, defining $m_{AB}=0$.  
In practice, this integration is performed by summing over the SDSS pixels which are given as $F_{\lambda}$ at intervals of $\Delta({\rm log}~\lambda)=0.0001$.  
If we sum over the pixels, then  
\begin{equation}
 g_{spec} = -2.5~{\rm log10}\, 
          \frac{\Sigma_i F_i\,w_i\,\lambda_i^2/c}
               {3631\times 10^{-23}~{\rm erg~s}^{-1}{\rm cm}^{-2}}
\end{equation}
where the summation is over all pixels (3800--$9200\rm\AA$).
and we have defined $w_i \equiv g_i/\Sigma_i g_i$.  

We then calculate the magnitude $g_{0spec}$ which would be measured in the $g$-band if the spectrum is de-redshifted by a factor $1+z$, back to $z=0$. To conserve flux, this `compression' increases the $F_{\lambda}$ normalization by a factor $1+z$, making 
\begin{equation}
 g_{0spec} = -2.5~{\rm log10}\, 
          \frac{(1+z)\Sigma_i F(\lambda_i/(1+z))\,w_i\,\lambda_i^2/c}
               {3631\times 10^{-23}~{\rm erg~s}^{-1}{\rm cm}^{-2}}.
\end{equation}
We define $r_{spec}$ and $r_{0spec}$ similarly, so the $k$-corrections 
are
\begin{equation}
 k_g = g_{0spec} - g_{spec} \qquad{\rm and}\qquad k_r = r_{0spec} - r_{spec}.
\end{equation}

\begin{figure}
\includegraphics[width=0.7\hsize,angle=-90]{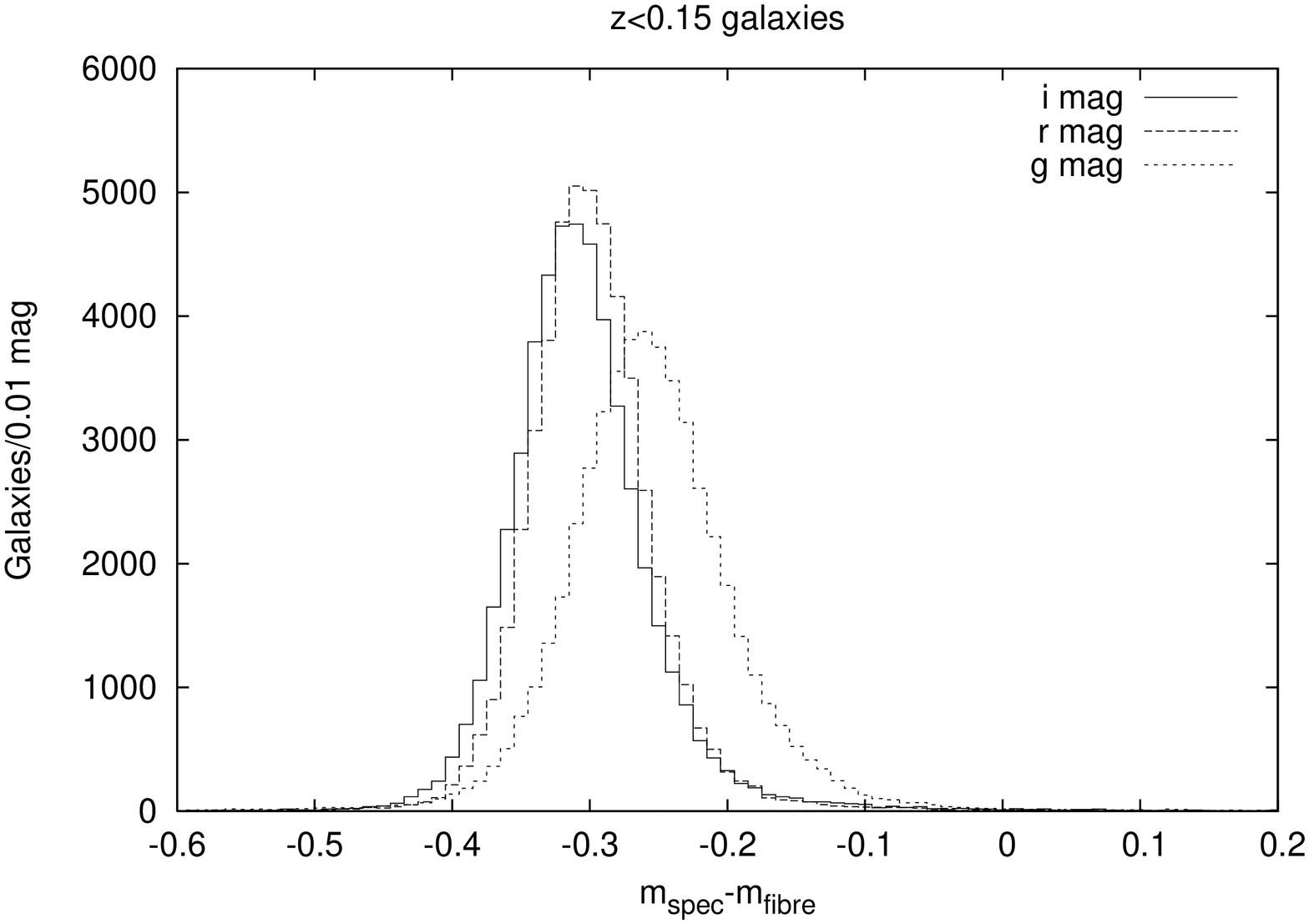}
\includegraphics[width=0.7\hsize,angle=-90]{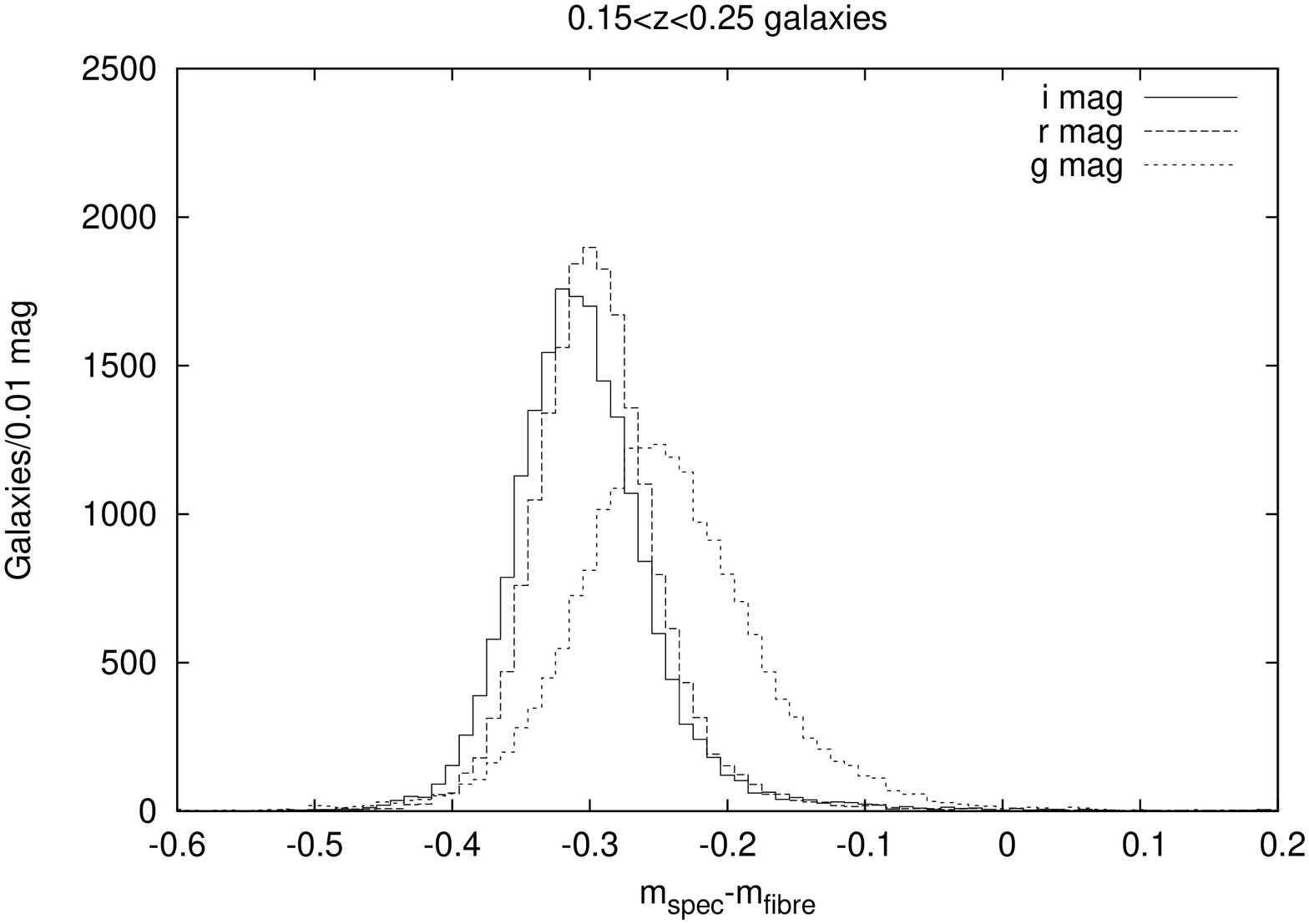} 
\includegraphics[width=0.7\hsize,angle=-90]{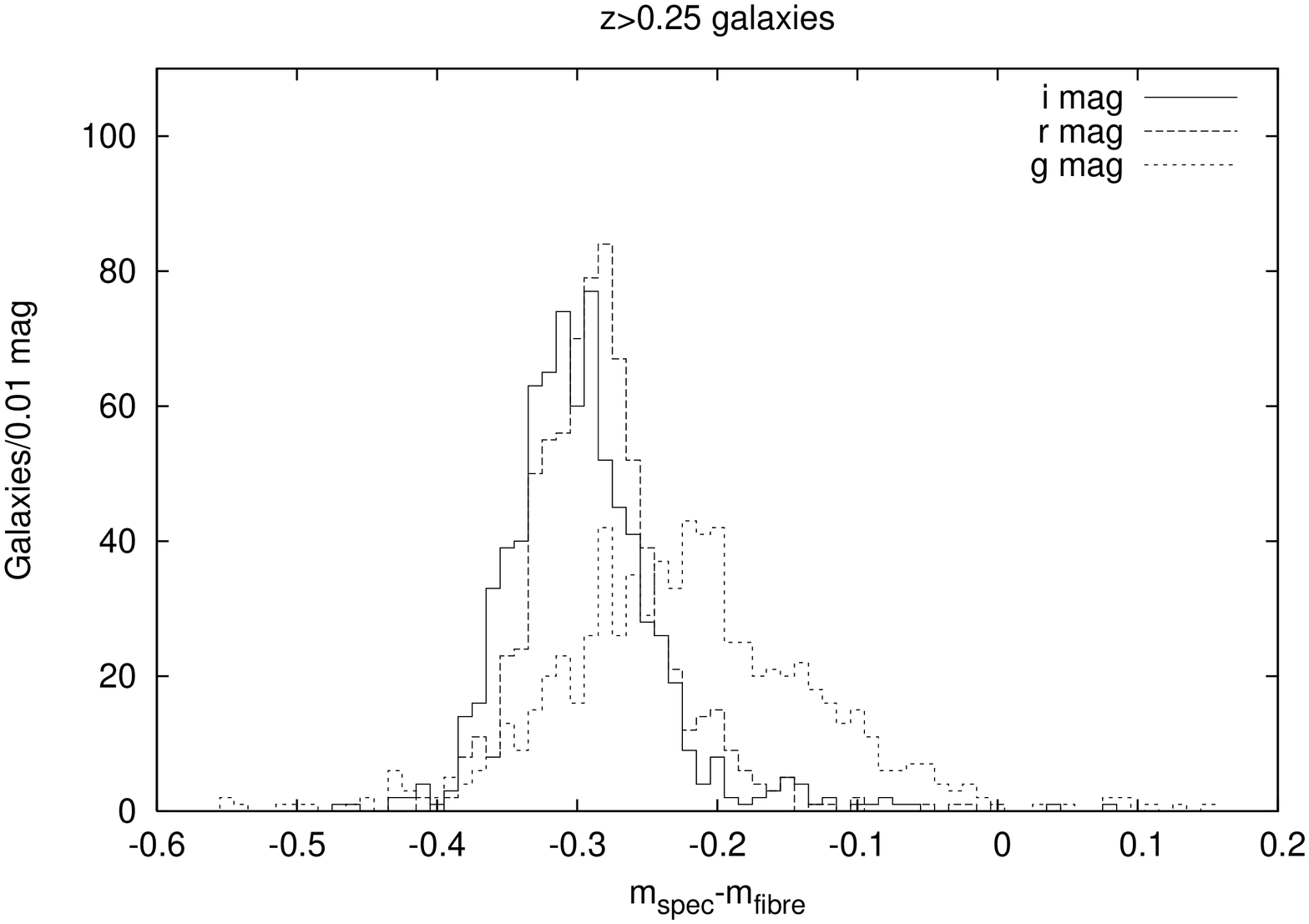}
\caption{Histogram of the difference between the spectra-derived and fiber (3 arcsec aperture) magnitude of SDSS early-type galaxies, plotted for the $g$, $r$ and $i$ bands and divided into 3 redshift intervals.}
 \label{sf-z}
\end{figure}

\begin{figure}
\includegraphics[width=0.7\hsize,angle=-90]{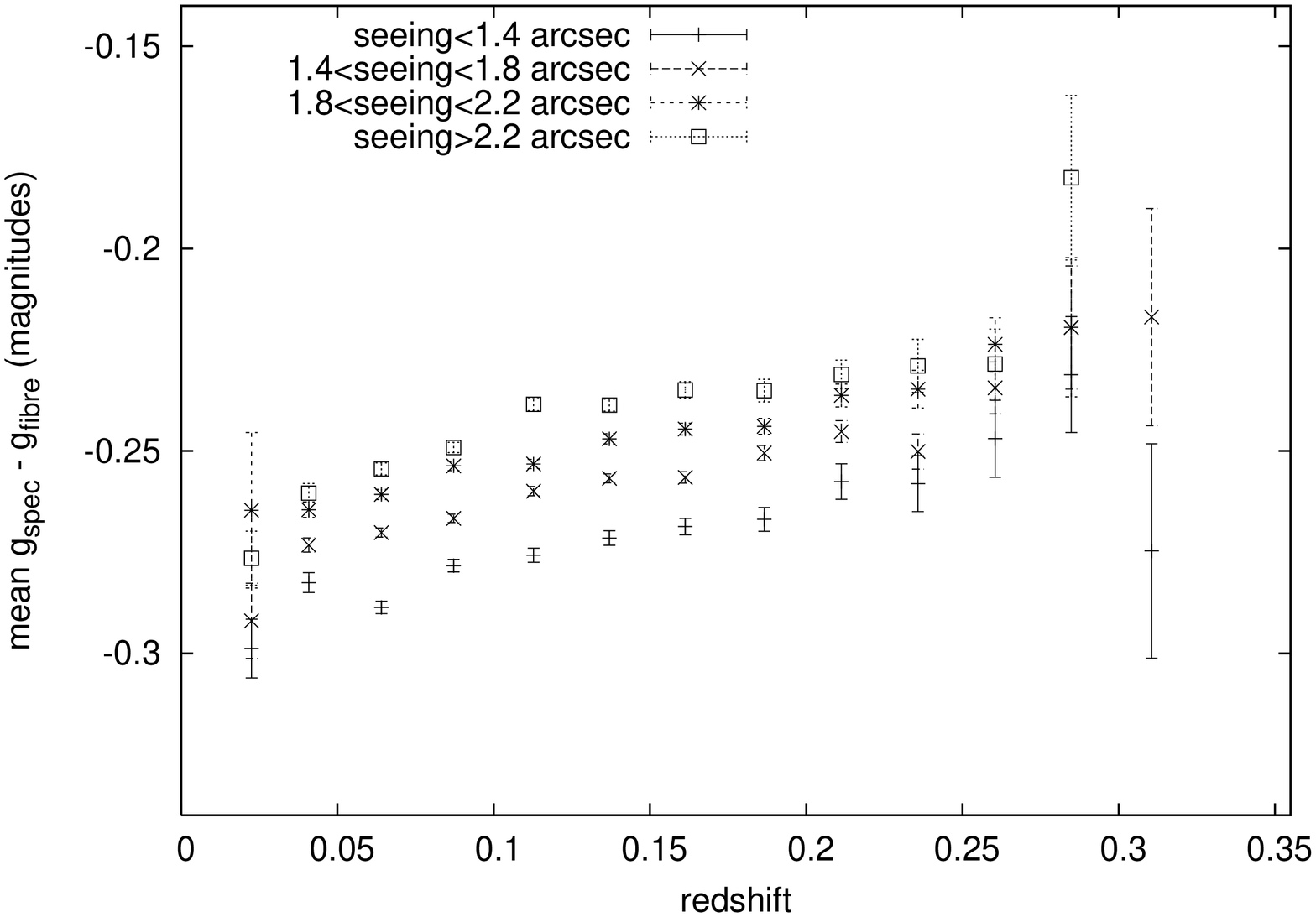}
\includegraphics[width=0.7\hsize,angle=-90]{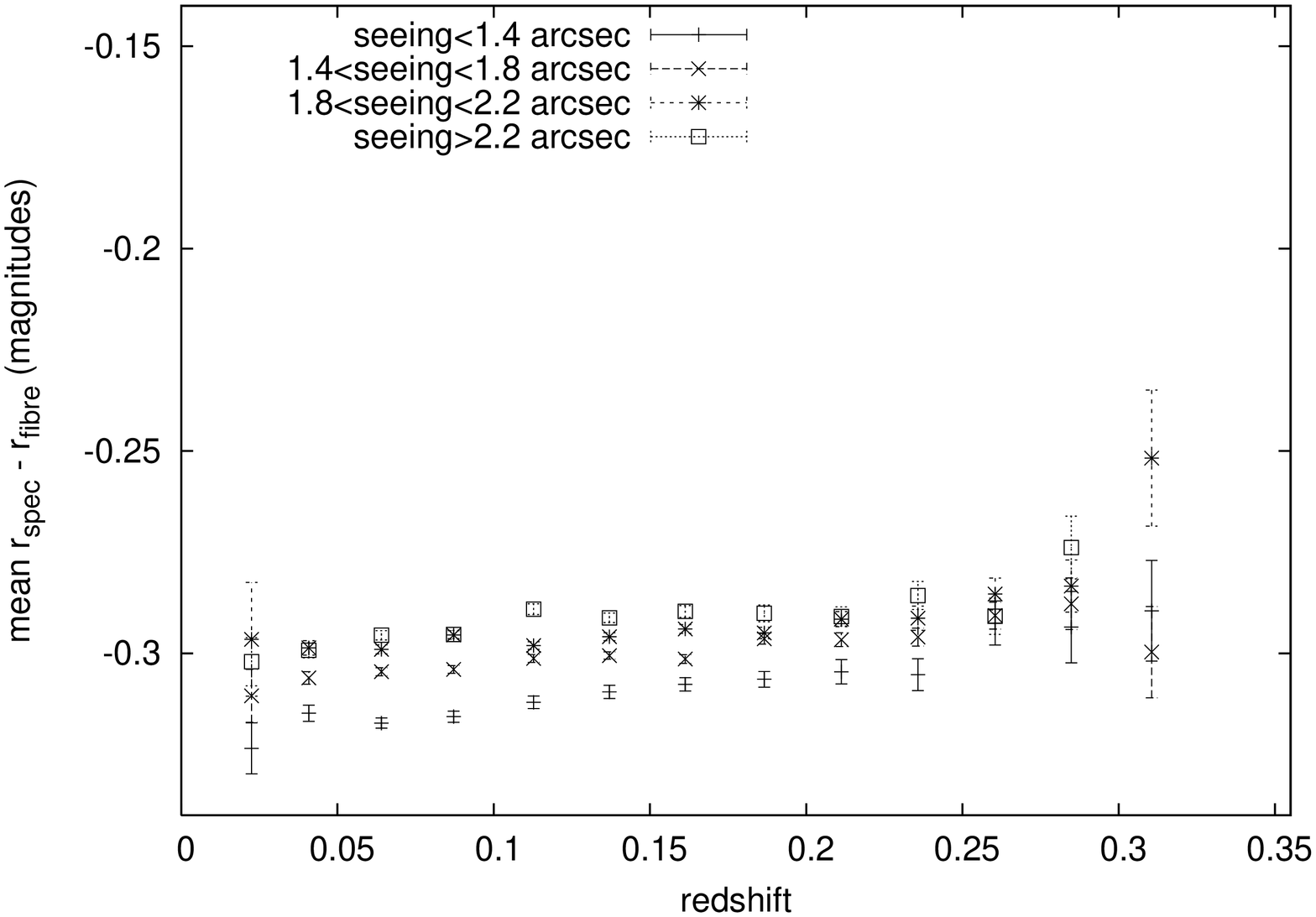} 
\caption{The mean difference between the spectra-derived and fiber (3 arcsec aperture) magnitudes of SDSS early-type galaxies, as a function of redshift with the galaxies divided into four intervals of the seeing during the spectroscopic observation. Shown for $g$ (top) and $r$ (bottom) bands.} 
 \label{sf-seeing}
\end{figure}

\subsection{Spectral and photometric calibrations}
Magnitudes derived from the spectra are compared with the (reddening-subtracted) {\tt fiber} (i.e. 3 arcsec aperture) magnitudes from the DR6 SDSS catalog. It is expected that these would be offset by $m_{spec} -m_{fib}\simeq -0.31$, because the spectra are flux-calibrated using the {\tt PSF} magnitudes of the spectrophotometric standard stars (Adelman-McCarthy et al. 2008), rather than the {\tt fiber} magnitudes (which are about 0.31~mags fainter).
 
Figure~\ref{sf-z} shows a histogram of $m_{spec}-m_{fib}$, in the $g$, $r$ and $i$ bands, for our E/S0 sample, in a few redshift bins. At $z<0.15$, we find $\langle r_{spec}-r_{fib}\rangle =-0.299$ and $\langle i_{spec}-i_{fib}\rangle=-0.308$, but $\langle g_{spec}-g_{fib}\rangle=-0.259$. This discrepancy means that the $g-r$ extracted from the spectra are on average 0.04 mag  redder than the fiber magnitude $g-r$. The calibration of the $g-r$ magnitudes from SDSS data is claimed to be accurate to $\simeq 0.02$ mag (Adelman-McCarthy et al. 2008), but it is conceivable the errors from the imaging and spectroscopic data might be in opposite directions from the true calibration, producing a 0.03--0.04 mag difference in their $g-r$. 

In addition, Figure~\ref{sf-seeing} shows that the difference between the flux recorded in the spectra and that on the imaging data is affected slightly by the atmospheric seeing during the spectroscopic observation. Imaging data are taken in good seeing, but $51\%$ of the spectra were taken in $r$-band seeing of $>1.8$ arcsec. We estimate each 1 arcsec increase in seeing produces a flux deficit
$\Delta(m_{spec}-m_{fib})$ of 0.021 mag in $g$ and 0.012 mag in $r$. Typical seeing could then redden the spectra by $\Delta(g-r)\simeq 0.01$ mag, but does not account for all the difference in spectra and fiber $g-r$. The relatively small effect of seeing on the spectrophotometry should not significantly affect our estimates of the CMR evolution as it does not appear to increase with redshift, or to fainter fluxes. 

\begin{figure}
\includegraphics[width=0.7\hsize,angle=-90]{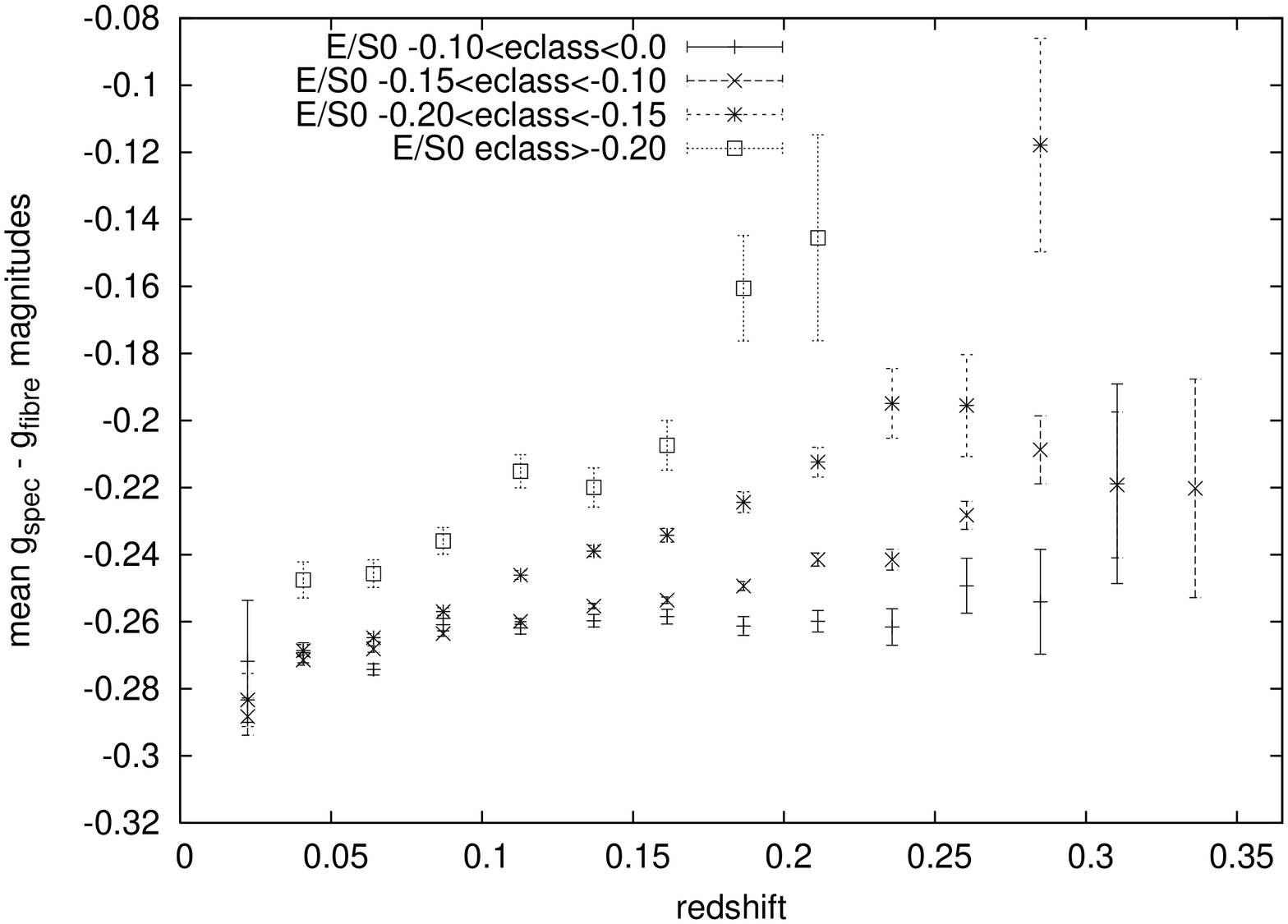}
\includegraphics[width=0.7\hsize,angle=-90]{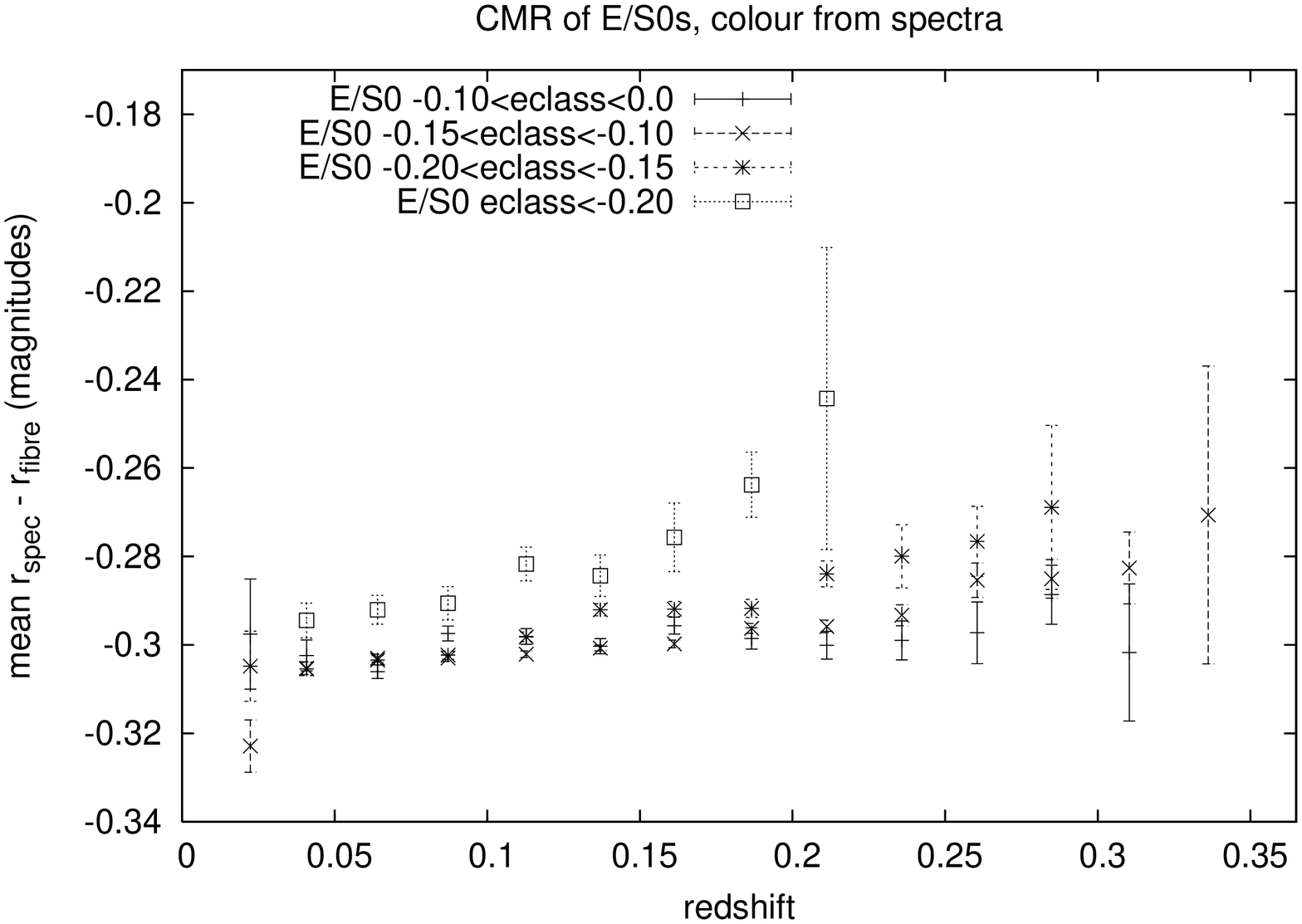}
\caption{The mean magnitude difference (a) $g_{spec}-g_{fiber}$ 
(b) $r_{spec}-r_{model}$ of SDSS early-type galaxies, as a function of redshift, divided into 4 intervals of the {\tt eclass} spectral classification parameter.} 
 \label{sf-eclass}
\end{figure}

However, Figures~\ref{sf-z} and~\ref{sf-seeing} also show that the mean $g_{spec}-g_{fib}$ increases with redshift, as does the scatter in this quantity.  This is not related to seeing but might be be a separate problem with the SDSS spectrophotometry. Further investigation of the $g_{spec}-g_{fib}$ offset found it was correlated with the {\tt eclass} spectral type classification parameter, being much greater for redder galaxies with more negative {\tt eclass} (Figure~\ref{sf-eclass}). Note the E/S0 sample all have $\rm eclass\leq 0$ with a mean of -0.1275. In the red-band the mean $r_{spec}-r_{fib}$ is also correlated with {\tt eclass}, although the effect is smaller and does not increase strongly with redshift. The dependence on {\tt eclass} is obviously non-linear and might be better described as {\tt eclass}$^2$.

\begin{figure}
\includegraphics[width=0.7\hsize,angle=-90]{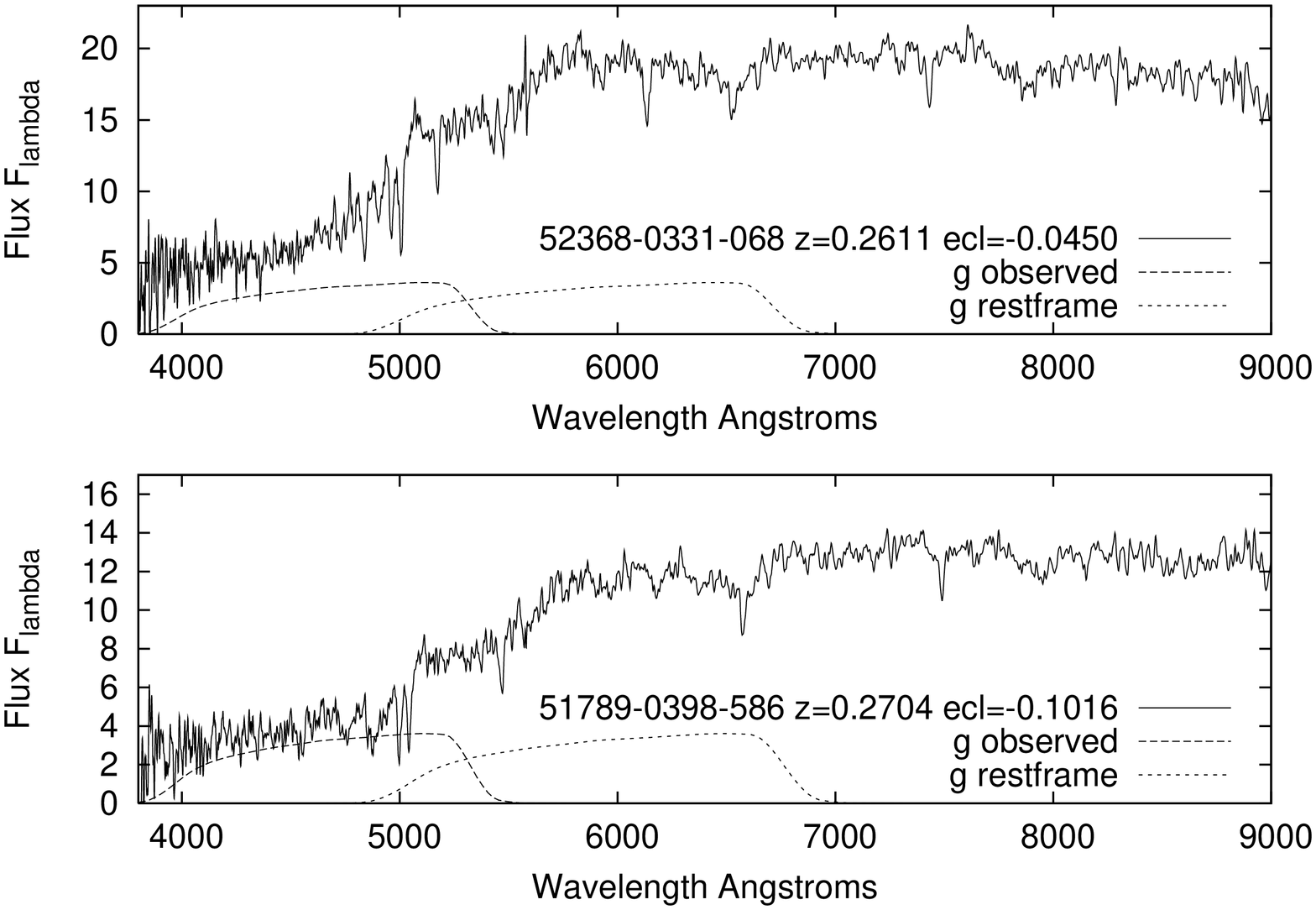}
\includegraphics[width=0.7\hsize,angle=-90]{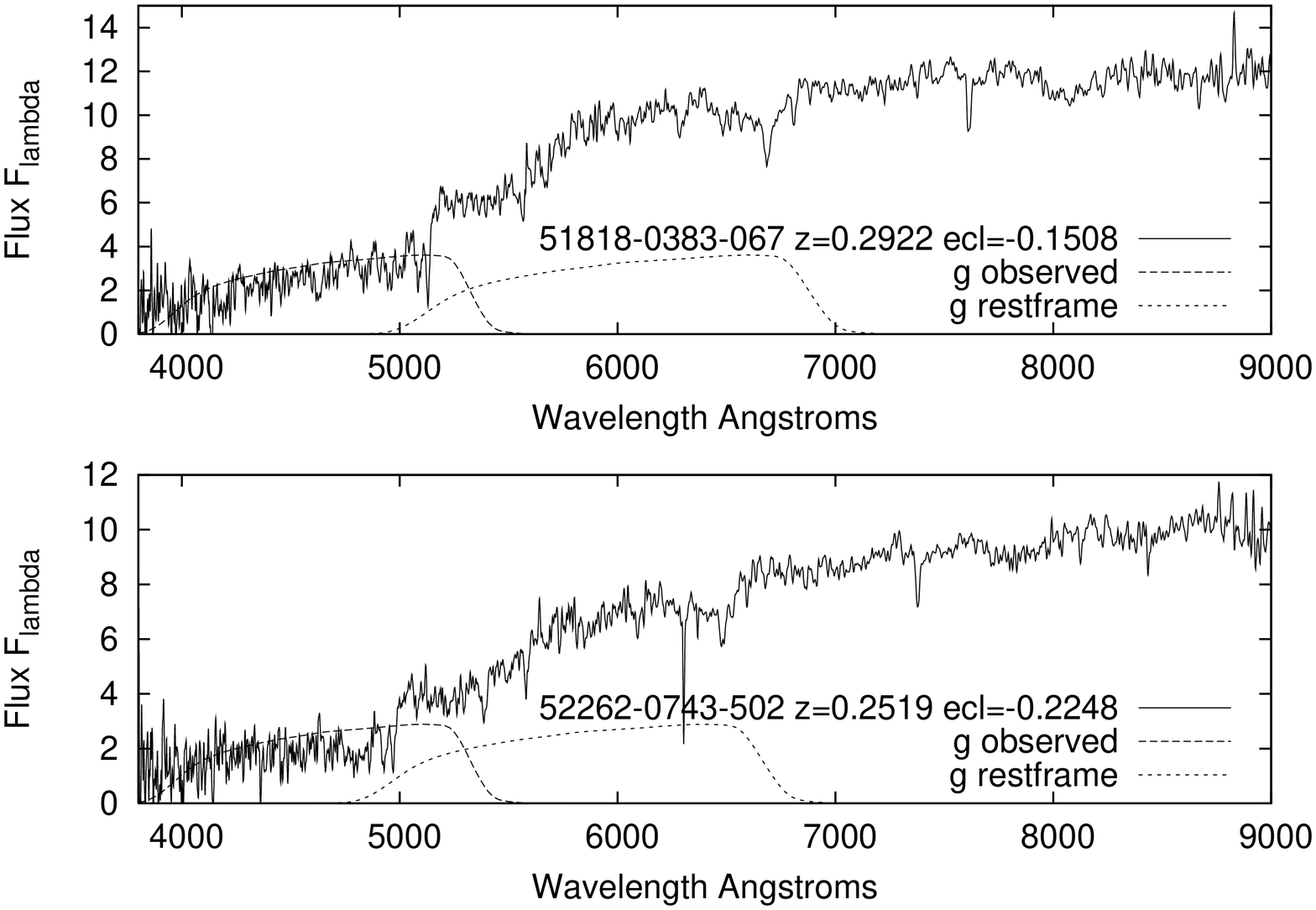}
\caption{Spectra of four representative $z\simeq 0.3$ E/S0s with similar magnitudes and a range of {\tt eclass}, showing the difference at the blue end, with the $g$-band response function in the observer frame and in the galaxy rest-frame. Spectra are smoothed (slightly) with $\sigma=2$ pixels.}
 \label{z0.3spec}
\end{figure}

To understand why $g_{spec}$ might be affected by {\tt eclass}, especially at higher redshifts,  we examine the spectra of four representative E/S0 galaxies of different {\tt eclass}, all at $z>0.25$. Figure~\ref{z0.3spec} illustrates that a more negative eclass is associated with a lower flux in the rest-frame UV, which at these redshifts is observed in the $g$-band. The spectra are particularly noisy at the blue end of the observed $g$-band, near $4000\rm \AA$. 

Flux variations from noise at the blue end of the spectrum would produce correlated variations in $g_{spec}$ and {\tt eclass}. But the trend in $\langle g_{spec}-g_{fib}\rangle $ with redshift implies that this scatter  is not symmetric, but skewed.  As a result, for spectra with very low signal/noise at these wavelengths, the integrated $g_{spec}$ flux is, on average, underestimated by a few times 0.01~mags.

\subsection{Empirically improved calibration}
\begin{figure}
\includegraphics[width=0.7\hsize,angle=-90]{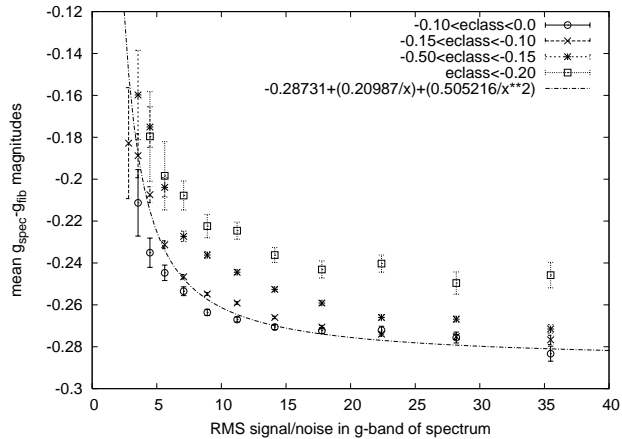}
\caption{The mean $g_{spec}-g_{fib}$ of early-type galaxies from the SDSS photometry, as a function of root-mean-square signal/noise $\rm pixel^{-1}$in the $g$-band of the spectrum. Galaxies are divided into 4 intervals of {\tt eclass}. A best-fit function of signal/noise ($x$) for the sample mean {\tt eclass} of -0.1275 is also plotted.}
 \label{sf-fit}
\end{figure}

Although the reason for this discrepancy in the SDSS spectrophotometry is not known, we can try to correct for it as follows.  We begin by assuming that the offset between the spectra-derived and fiber magnitude is not directly sensitive to redshift or wavelength, but depends only on:\\
  (i) the signal/noise in the spectrum, averaged across the passband,\\
 (ii) the tilt of the spectrum across the band, represented by 
      {\tt eclass}$^2$,\\
(iii) a cross-term ({\tt eclass}$^2\times$ signal/noise). \\
For each galaxy we calculate directly from the SDSS spectrum and the associated noise spectrum, the root-mean-square signal/noise per pixel in 4 wavelength intervals, the observer-frame and the rest-frame $g$ and $r$ bands.  Figure~\ref{sf-fit} shows  $\langle g_{spec}-g_{fib}\rangle$ as a function of rms signal/noise, for four bins in {\tt eclass}.   The mean trend is well described by 
\begin{eqnarray}
 \langle g_{spec}-g_{fib}\rangle &\approx&
   -0.28731 - \frac{0.09186}{x} + \frac{0.505216}{x^2} \nonumber\\
                    && \qquad   - 0.32549 y - 7.2538 \frac{y}{x},
 \label{eq-corr}
\end{eqnarray}
where $x$ is the rms signal/noise and $y$ is {\tt eclass}$^2$. The rms residual is 0.060 magnitudes.  The smooth curve in the Figure shows this expression when {\tt eclass}$=-0.1275$.  

We define our signal/noise correction function to be the final four terms in equation~(\ref{eq-corr}),

\begin{figure}
\includegraphics[width=0.7\hsize,angle=-90]{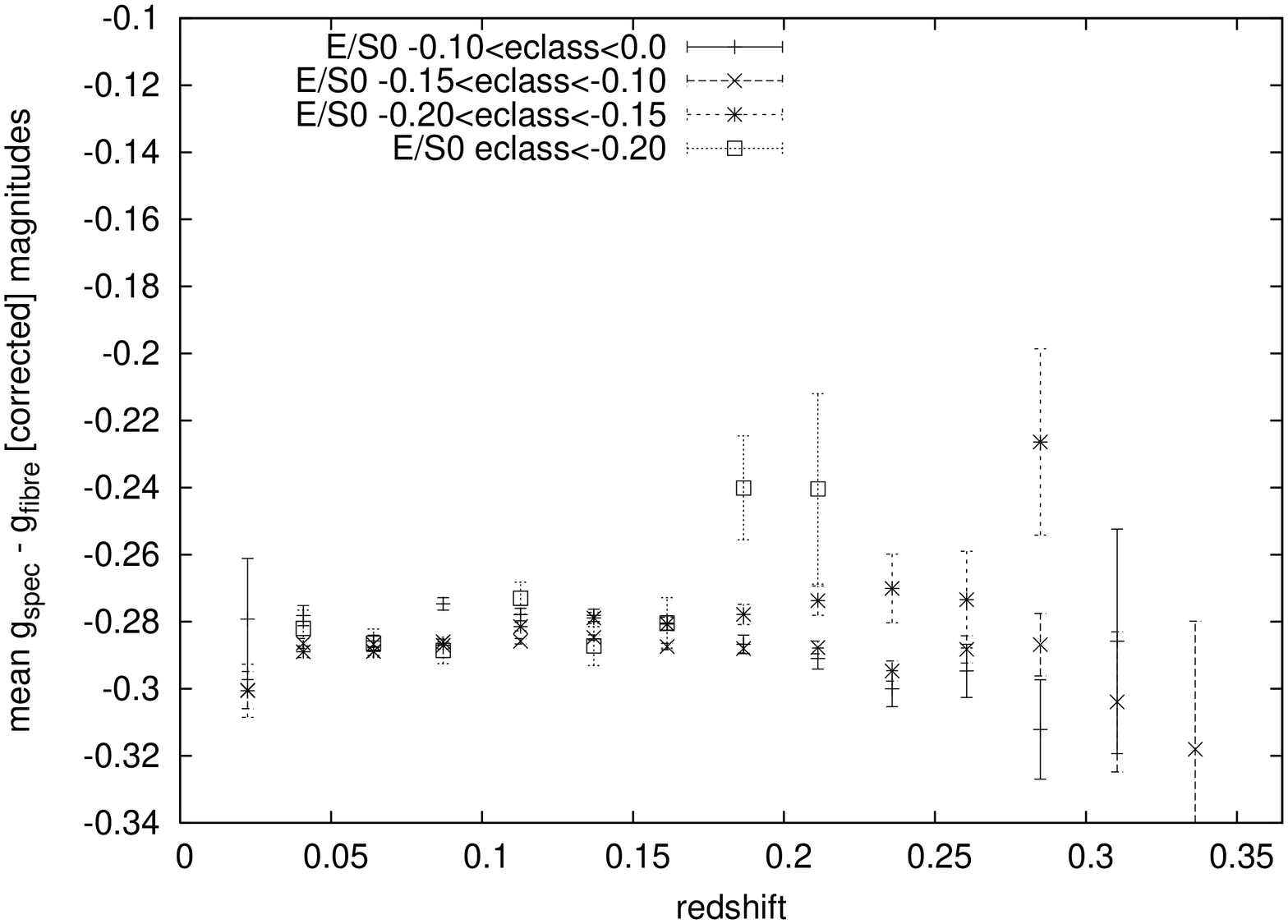}
\includegraphics[width=0.7\hsize,angle=-90]{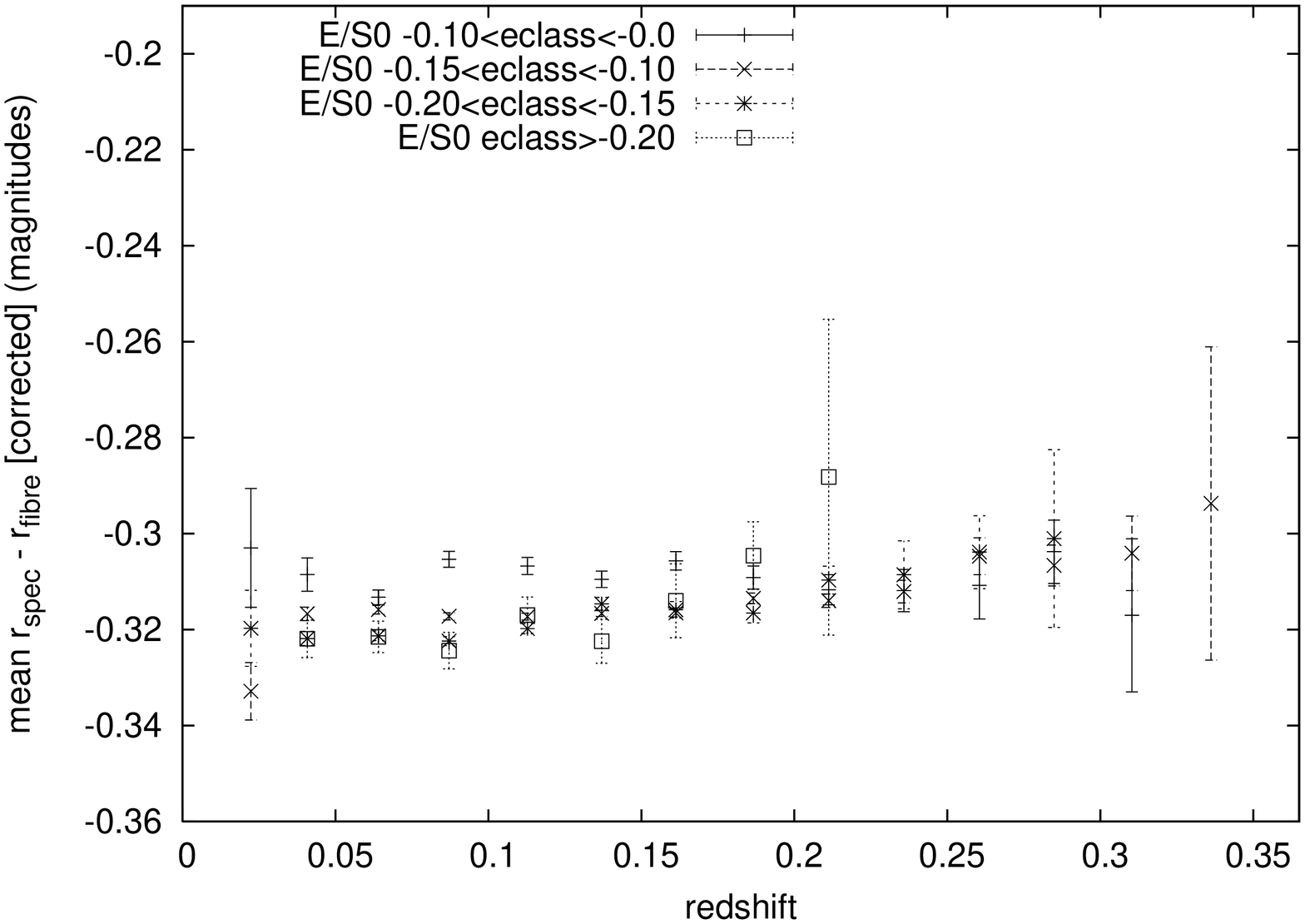}
\caption{The mean magnitude difference (a) $g_{spec}-g_{fiber}$ 
(b) $r_{spec}-r_{model}$ of SDSS early-type galaxies, as a function of redshift, divided into 5 intervals of {\tt eclass}, and using the 'corrected'
$g_{spec}$ and $r_{spec}$ magnitudes.} 
 \label{sf-corr}
\end{figure}

Before proceeding further, we illustrate the effectiveness of this correction in removing the differences between the spectra and image photometry, in particular the redshift trend in Figure~\ref{sf-eclass}.  Figure~\ref{sf-corr} is similar to Figure~\ref{sf-eclass}, except that now the signal/noise corrections (based on equation~\ref{eq-corr}) have been subtracted from both $g_{spec}$ and $r_{spec}$.  Almost all the redshift and {\tt eclass} dependence has now been removed from $g_{spec}-g_{fib}$, although there may be slight residuals especially at $z>0.25$.  
Similarly, by using the rms signal/noise in the rest-frame $g$ and $r$, we obtain the corrections for $g_{spec0}$ and $r_{spec0}$, and then a set of `corrected' k-corrections for each galaxy from the $g_{spec}-g_{spec0}$ and $r_{spec}-r_{spec0}$. 

\subsection{Comparison with models}
We now compare our estimated $k$-corrections with those derived from models which represent the range of expected ages and star-formation histories for E/S0s (e.g. Jimenez et al. 2007).  We do this by using the  Bruzual \& Charlot (2003) spectrophotometric models, with the (recommended) Padova 1994 stellar evolution tracks, the Chabrier (2003) IMF and fixed solar metallicity.

In our Model 1, star-formation (SF) begins 12.5~Gyr ago ($z=4.53$), continues at a constant rate for 1~Gyr until 11.5~Gyr ago ($z=2.83$) and then ceases. In Models 2, 3 and 4, SF begins 12.0~Gyr ago at ($z=3.49$) and the SF rate decreases exponentially with timescales $\tau=1$, 1.5 and 2~Gyr. The models include no internal dust extinction after the cessation of SF. Over $0<z<0.3$, models (1,2,3,4) predict mean rates of
 luminosity evolution $\Delta(M_r)=-(1.11, 1.20, 1.37, 1.52)z$ and
 rest-frame colour evolution $\Delta(g-r)=-(0.124, 0.171, 0.265, 0.366)z$.

We can also estimate the evolution of the luminosity function characteristic magnitude $M^*_r$ directly from the data using a $V/V_{max}$ method. We sum $V/V_{max}$ with the rest-frame absolute magnitudes adjusted faintward by a linear function of redshift $+Qz$, which is a reasonable approximation at these moderate redshifts. The mean $V/V_{max}$ matches the no-evolution value of 0.50 for $Q=0.86$, thus giving an estimate of the luminosity function evolution as $\Delta(M_r)\simeq -0.86z$. This agrees with the $-0.85z$ estimate of Bernardi et al. (2003b) for SDSS early-types and the $-0.90z$ of Hyde \& Bernardi (2009). The difference from the slightly stronger evolution in the BC03 models could be explained by a small amount of merging.

\begin{figure}
\includegraphics[width=0.7\hsize,angle=-90]{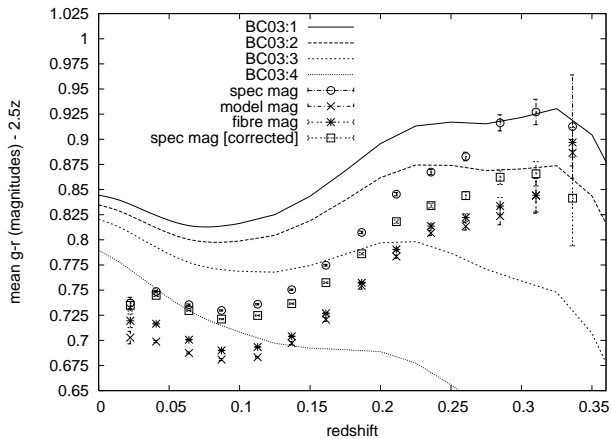}
\caption{Mean observer-frame $g-r$ colours of the E/S0 galaxies as a function of redshift, calculated from the spectra -- before and after the signal/noise `correction' -- and from the {\tt fiber} and the {\tt model} magnitudes. These $g-r$ colours are compared with four evolving BC03 models for early-type galaxies (see text). For clarity a linear trend $2.5z$ is subtracted in all cases.}
 \label{bc03colours}
\end{figure}

Figure~\ref{bc03colours} shows the mean observer-frame $g-r$ colours (spectra-derived, {\tt model} and {\tt fiber}) of our E/S0 sample as a function of redshift, compared to the BC03 models. No single model fits over the whole redshift range, because with increasing redshift, more intrinsically luminous galaxies dominate and these (as we shall see) are significantly redder. The uncorrected spectra-derived  colours gradually cross from those of BC03 model 4 at $z\sim 0$ to the redder model 1 at $z>0.3$, while the mean {\tt fiber} and {\tt model} magnitude colours are at all $z$ a few 0.01 mag bluer, and closer to model 2 at $z>0.3$. The signal/noise corrected colours are intermediate, being similar to the uncorrected colours at low redshift but meeting model 2 at $z=0.3$. The mean {\tt model} colours are  slightly ($<0.02$ mag) bluer than the {\tt fiber} colours, especially at $z<0.1$,  which may be the result of radial colour gradients.

\begin{figure}
\includegraphics[width=0.7\hsize,angle=-90]{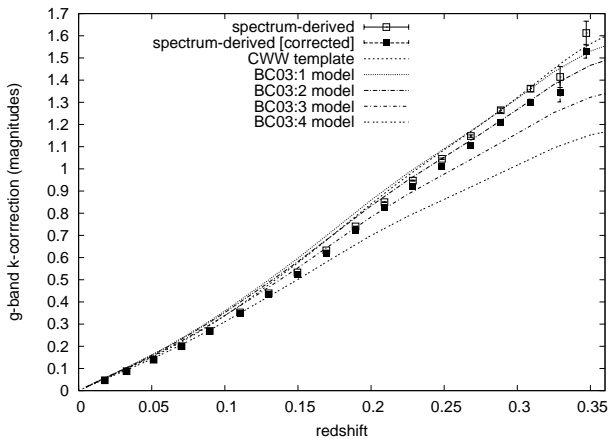}
\includegraphics[width=0.7\hsize,angle=-90]{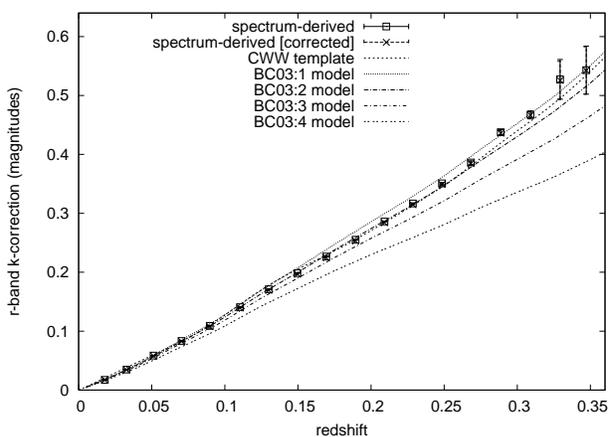} 
\caption{Mean k-corrections for the E/S0 sample, $k_g$ and $k_r$, calculated from both uncorrected and corrected spectra magnitudes, compared with four evolving BC03 models and also and the $k$-correction from the CWW non-evolving template spectrum for E/S0s.} 
 \label{bc03kcorr}
\end{figure}

Figure~\ref{bc03kcorr} shows the spectra-derived k-corrections of the E/S0s -- from the uncorrected and corrected magnitudes -- averaged in $\Delta(z)=0.025$ intervals, and compared with the four BC03 models and CWW template spectra. The k-corrections are tabulated in Tables 1\&2. Note that at redshift $z$, the BC03 model k-correction depicted is that calculated on the model spectrum at the lookback time $t(z)$, and not the k-correction of the $z=0$ spectrum, i.e. we plot the quantity $M_{rf}(z)-M_{ev}(z)$, and not `$k$' from the `magnitude\_FN' file (see BC03).

The signal/noise correction reduces the $g$-band k-correction ($k_g$) at higher redshifts, by $\simeq 0.05$ magnitudes at $z=0.3$, but the reduction in $k_r$ remains very small at $<0.01$ magnitudes. The mean $k$-corrections from the spectra are generally in the range of the BC03 model predictions, nearest to Model 3 or 4 at lower redshift and the redder Models 1 (uncorrected) or 2 (corrected) at $z>0.2$. 

\section{Colour-magnitude Relation (CMR)}
We now study the colour-magnitude relation and its evolution.
In all cases, the absolute magnitude $M_r$ we use is 
\begin{equation}
 M_r = r - A_r - k_r - d_{mod}(z) + Qz,
\end{equation}
where $r$ is the {\tt model} $r$ magnitude from the photometry, 
 $A_r$ is the Galactic reddening in the $r$-band at the sky position of 
 the galaxy (as given in the DR6 catalog),
 $k_r$ is the $r$-band k-correction, and
 $d_{mod}(z)$ the distance modulus at the spectroscopic redshift $z$.  
We discuss the effects of different choices for the $k$-correction below.  
The final term $Qz$, with $Q=0.86$ here, `de-evolves' the absolute magnitudes to $z=0$, so as to isolate the effect of colour evolution from that of luminosity evolution.  
%, which because of the slope of the CMR would otherwise make a (small) contribution to its blueward shift. 

The rest-frame colour, 
\begin{eqnarray}
 (g-r)_{rf} &=& g - A_g - k_g - (r - A_r - k_r) \nonumber\\
            &=& (g-r) - (A_g - A_r) - (k_g - k_r),
\end{eqnarray}
can be calculated in a number of different ways.  If we use the spectra to estimate $(g_0-r_0) = (g-r)-(k_g-k_r)$, then these are colours within a fixed angular aperture.  We compare this with the result of using $g-r$ from the {\tt fiber} photometry, but $k_g-k_r$ from the spectra.  We then show what happens if we determine the $k$-corrections directly from the {\tt fiber} colours.  

To explore the dependence on aperture, we also use {\tt model} $g-r$ from the photometry, which gives colours within a physical scale which is related to the $r-$band half-light radius of each galaxy.  In this case, if we determine the $k$-correction from the spectra, then we have colours from one scale in the galaxy, but $k$-corrections potentially from another.  Therefore, we also show the result of deriving $k$-corrections from the {\tt model} photometry, as was done for the {\tt fiber} photometry.

In all cases, we did not find evidence for evolution in CMR slope with redshift. On the basis of this, and previous studies finding no CMR slope evolution even to $z\sim 1.3$ (e.g. Mei et al. 2009), we fit (using {\sevensize IRAF} {\tt surfit}) 
\begin{equation}
 \Bigl\langle (g-r)_{rf}|M_r,z\Bigr\rangle = a_0 + a_1\,M_r +a_2\,z,
 \label{cmrfit}
\end{equation}
with separate linear dependencies on $M_r$ and $z$ and no cross-term.  A few galaxies with outlying colours $(g-r)_{rf}<0.5$ or $>1.1$ were excluded from the slope and evolution fits ($0.55\%$ of the total, leaving 69987 objects).

\begin{figure} 
\includegraphics[width=0.7\hsize,angle=-90]{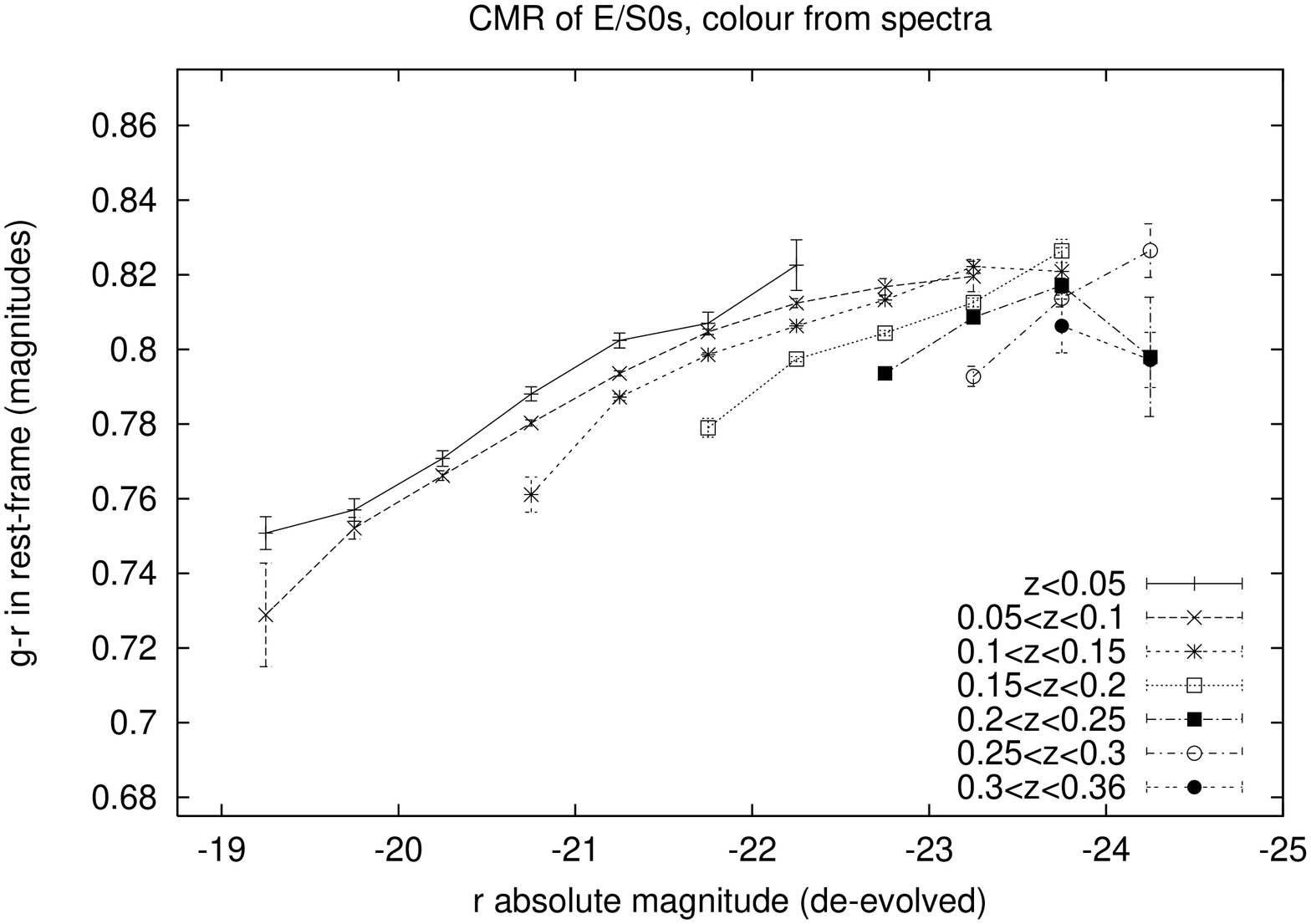} 
\includegraphics[width=0.7\hsize,angle=-90]{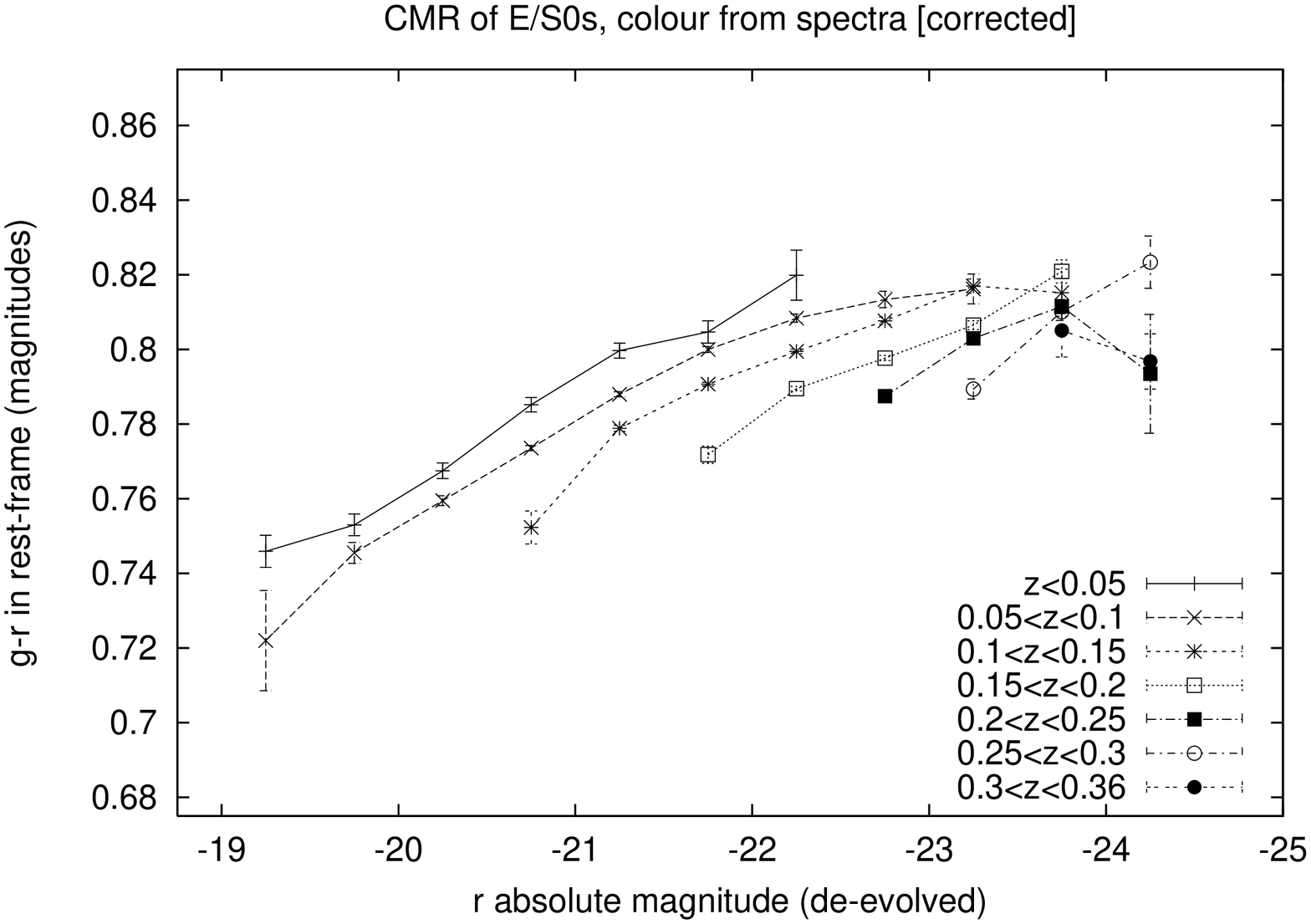}
\caption{Colour-magnitude relation ($g-r$ against $M_r$ in the rest-frame) 
for SDSS E/S0 galaxies in seven $\Delta(z)=0.05$ intervals of redshift ($0.30<z<0.36$ for the seventh), calculated using rest-frame $g$ and $r$ derived directly from the SDSS spectra, before (a) and after (b) applying the signal/noise correction (equation~\ref{eq-corr} and associated discussion).  For clarity, points corresponding to fewer than 10 galaxies are not plotted.}
\label{cmr-spec} 
\end{figure}

\subsection{CMR for a fixed angular scale}
Figure~\ref{cmr-spec} shows the CMR from $(g-r)_{rf}$ colours directly measured from the spectra, with and without the signal/noise correction. The galaxies are divided into 7 redshift intervals to illustrate the evolution. The signal/noise correction has much less effect on the rest-frame $g_{spec}$ than on the observed-frame  $g_{spec}$, and consequently it produces much less shift in the spectra-derived CMR than the $g$-band k-correction. The colours of higher redshift galaxies are shifted bluewards, but by no more than 0.01 mag.
%Figure~\ref{cmr-spec} shows the CMR for colours derived from the spectra, before and after applying the corrections associated with equation~(\ref{eq-corr}).  
The CMR, before and after correction, is well described by equation~(\ref{cmrfit}) with 
% $\langle (g-r)_{rf}|M_r,z\rangle=(0.2825\pm 0.0086) 
%-(0.02486\pm 0.00042) M_r -(0.2368 \pm 0.0069)z$
\begin{eqnarray}
 a_0 &=& (0.2825\pm 0.0086) \quad{\rm or}\quad (0.2426\pm 0.0083) \nonumber\\
 a_1 &=& -(0.02486\pm 0.00042) \quad{\rm or}\quad -(0.02659\pm 0.00041)\nonumber \\
 a_2 &=& -(0.2368 \pm 0.0069) \quad{\rm or}\quad  -(0.2709 \pm 0.0067)\nonumber
\end{eqnarray}
with an rms scatter of 0.0541 magnitudes (Figure~\ref{cmr-spec}).
Notice that the correction slightly increases the CMR evolution;
% from $\Delta(g-r)=-(0.22-0.24)z$ to $-(0.26-0.27)z$;
it has little effect on the slope.

On the other hand, the CMR appears slightly curved.  This might cause a linear fit to mis-estimate the colour evolution.  Fitting a quadratic 
\begin{equation}
 \Bigl\langle (g-r)_{rf}|M_r,z\Bigr\rangle = a_0 + a_1\,M_r + a_2\,M_r^2+a_3\,z,
\end{equation}
yields
\begin{eqnarray}
 a_0 &=& (0.5119\pm 0.1066) \quad{\rm or}\quad (0.1722\pm 0.1035) \nonumber\\
 a_1 &=& -(0.0985\pm 0.0099) \quad{\rm or}\quad -(0.0650\pm 0.0096)\nonumber \\
 a_2 &=& -(0.00171 \pm 0.00023) \quad{\rm or}\quad  -(0.00089 \pm 0.0022)\nonumber \\
a_3 &=& -(0.2227 \pm 0.0072) \quad{\rm or}\quad  -(0.2635 \pm 0.0069)\nonumber 
\end{eqnarray}
This reduces the evolution only slightly.

\begin{figure} 
\includegraphics[width=0.7\hsize,angle=-90]{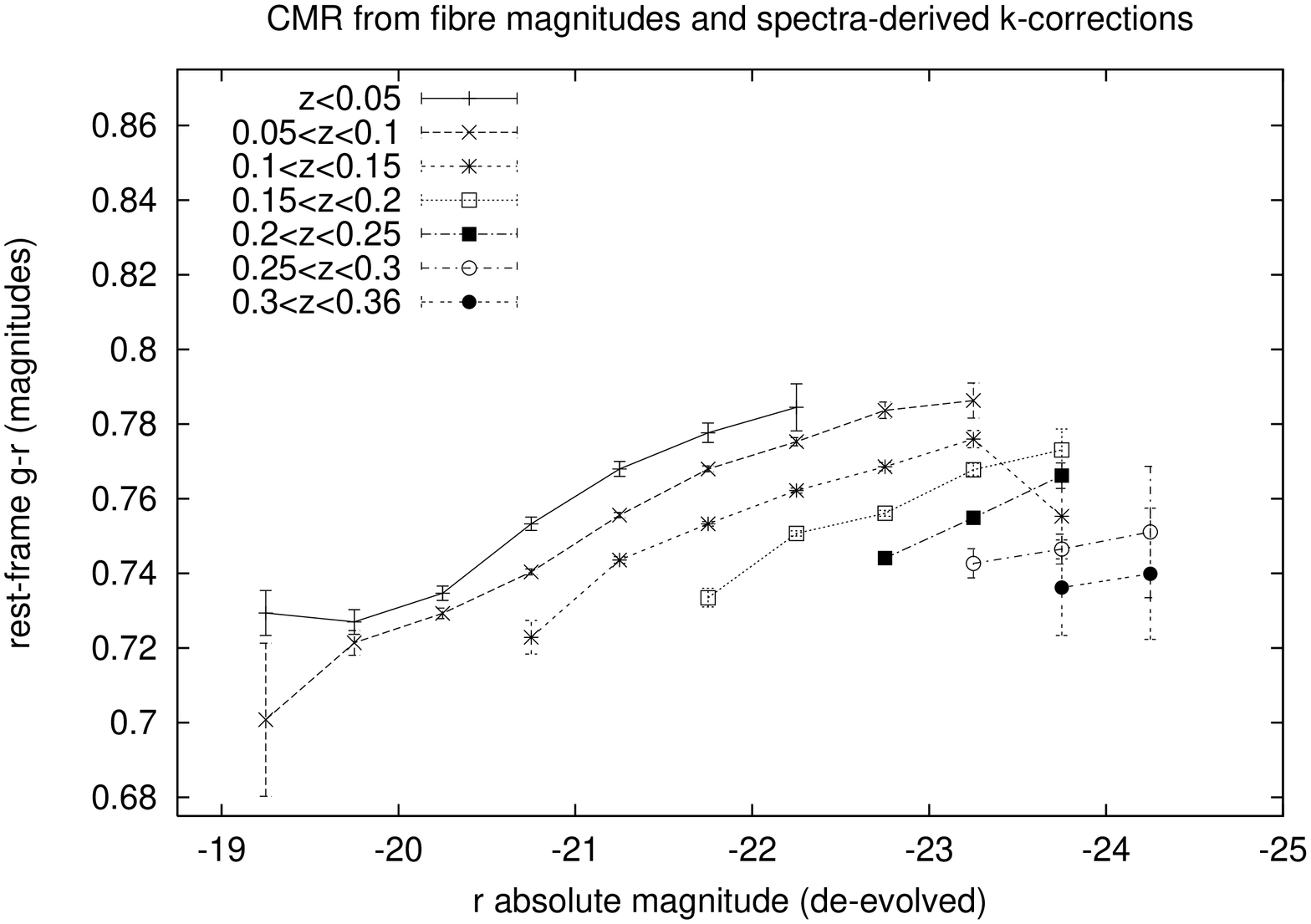}
\includegraphics[width=0.7\hsize,angle=-90]{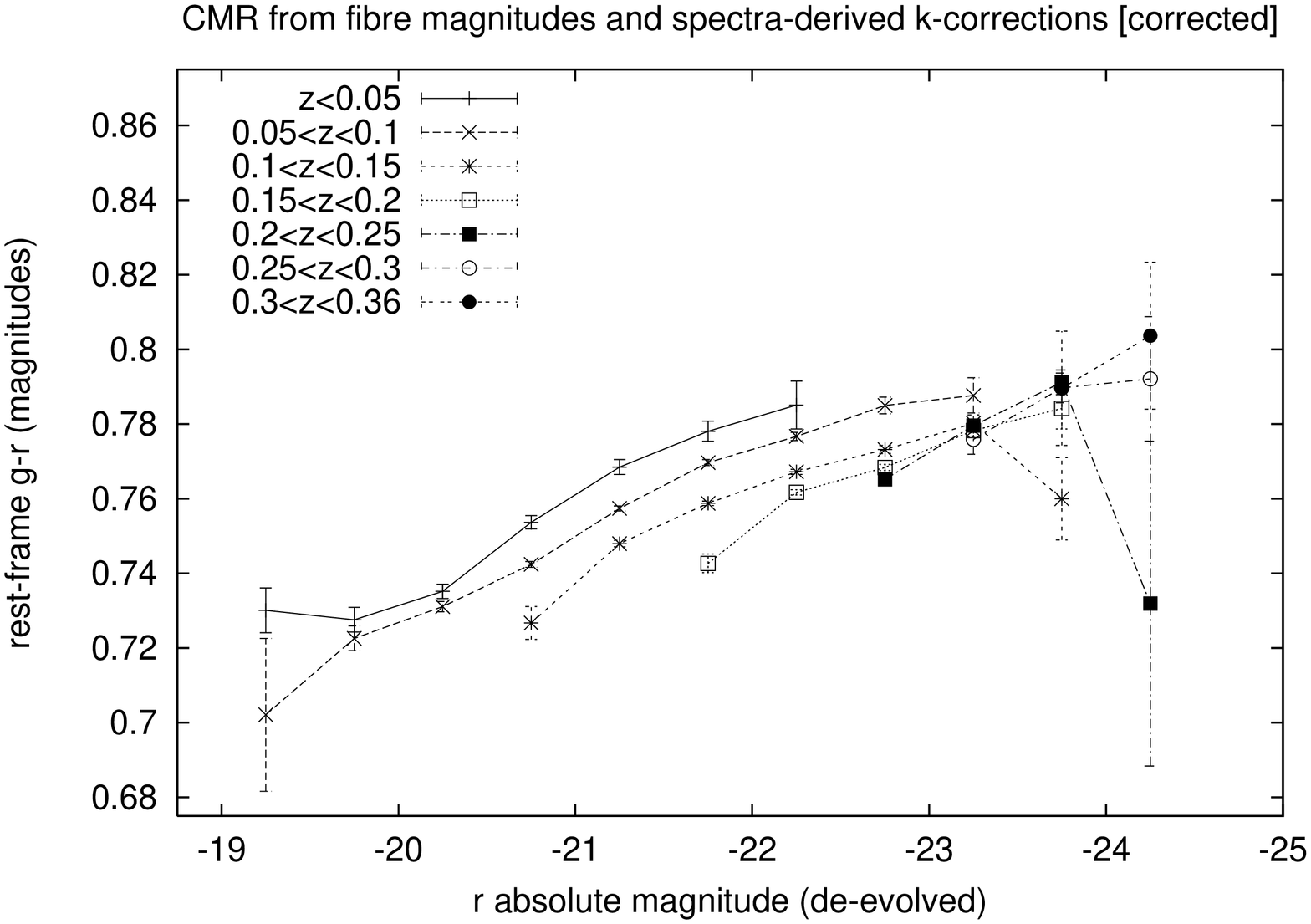} 
\caption{Same as previous figure, but now $g-r$ is from observed {\tt fiber} magnitudes, shifted to rest-frame using k-corrections derived from the spectra before (a) and after (b) applying the signal/noise correction.} 
 \label{cmr-fiber}
\end{figure}

Figure~\ref{cmr-fiber} shows the CMR from {\tt fiber} (3 arcsec aperture) magnitudes, corrected to rest-frame using the $k$-corrections from the spectra.  In this case, applying the signal/noise based correction makes a big difference.  We find 
\begin{eqnarray}
 a_0 &=& (0.2465\pm 0.0092) \quad{\rm or}\quad (0.2685\pm 0.0094)\nonumber\\
 a_1 &=& -(0.02517\pm 0.00045) \quad{\rm or}\quad -(0.02363\pm 0.00046)\nonumber\\
 a_2 &=& -(0.3480 \pm 0.0074) \quad{\rm or}\quad -(0.1999 \pm 0.0075)\nonumber
\end{eqnarray}
before and after the correction.  The uncorrected $k_g$ and $k_r$ yields very strong evolution; the correction reduces this substantially (about $0.15z$).

\subsection{CMR for a `fixed' physical scale}
\begin{figure} 
\includegraphics[width=0.7\hsize,angle=-90]{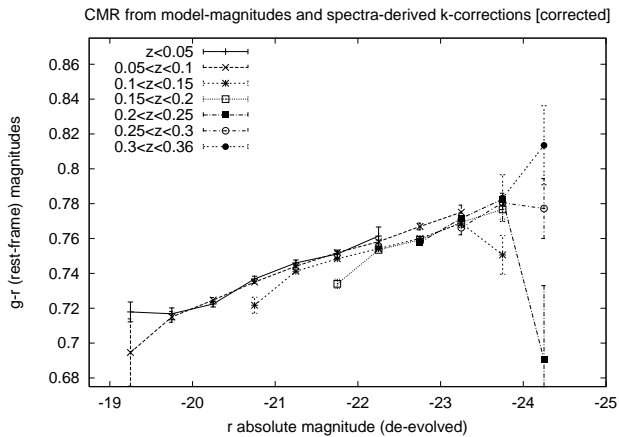} 
\caption{Rest-frame colour-magnitude relation in which $g-r$ is based on {\tt model} colours from the photometry and (signal/noise corrected) $k$-corrections from the spectra. }
 \label{cmr-model}
\end{figure}

Figure~\ref{cmr-model} shows the CMR obtained using the observed {\tt model} colours corrected to rest-frame with the spectra-derived k-corrections.  This relation is well fit by 
\begin{eqnarray}
 a_0 &=& (0.3543\pm 0.0091) \quad{\rm or}\quad (0.3794\pm 0.0093) \nonumber\\
 a_1 &=& -(0.01904\pm 0.00045) \quad{\rm or}\quad -(0.01735\pm 0.00046)\nonumber \\
 a_2 &=& -(0.2242 \pm 0.0074) \quad{\rm or}\quad -(0.0743 \pm 0.0075)\nonumber
\end{eqnarray}
with an rms scatter of 0.0578 magnitudes. 
Notice that the rest-frame colours are $\sim 0.04$ mag bluer than before, and the evolution is much smaller when the signal/noise $k$-corrections are used.  In addition, the CMR is shallower and less curved.

There is nothing physically significant about the size of the SDSS fiber, but the half-light radius may be a physically meaningful scale -- perhaps this is why it shows less curvature than the CMRs based on a fixed angular aperture. The fact that this CMR is bluer than that associated with a 3 arcsec aperture is due to colour gradients:  the typical half-light radius of the objects in our sample is larger than 3 arcsec, and E/S0s tend to be redder in the center.  Many previous CMR studies e.g. van Dokkum et al  (2000) and Mei et al. (2006, 2009) used colours measured in half-light apertures, so as to minimize colour gradient effects. The fact that the slope becomes shallower for the {\tt model} colours may have implications for how gradients scale with luminosity, which we return to shortly.  However, the fact that the evolution in this case is smaller than for a fixed angular aperture indicates that gradients have contributed to our other two estimates of the evolution of the CMR.  (A fixed angular scale collects light from a larger physical radius, where the colours are expected to be bluer even if there were no evolution, for the high redshift objects).

  Thus, there are many senses in which this CMR is the one with the most direct physical information.  Unfortunately, it is based on $k$-corrections which are computed for a different physical scale than that on which the colour itself is measured.  The next subsection studies the magnitude of this effect.  

\begin{figure} 
\includegraphics[width=0.7\hsize,angle=-90]{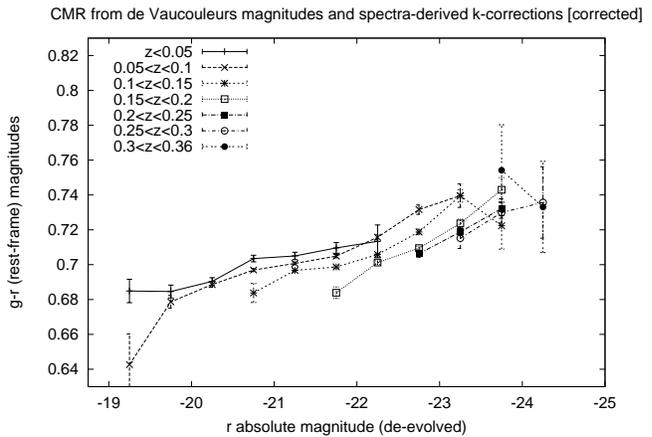} 
\caption{Rest-frame colour-magnitude relation in which $g-r$ is based on de Vaucouleurs profile fits to the galaxies in $g$ and $r$ and (signal/noise corrected) $k$-corrections from the spectra. }
 \label{cmr-dev}
\end{figure}

 Yet another measure of `total' magnitudes and colours may be obtained by fitting de Vaucouleurs profiles independently in $g$ and $r$, without constraining the effective radii to be the same. Figure~\ref{cmr-dev} shows the CMR from this type of photometry (with the signal/noise correction). This can be fitted with
\begin{eqnarray}
 a_0 &=& (0.3455\pm 0.0121) \nonumber\\
 a_1 &=& -(0.01726\pm 0.00060) \nonumber \\
 a_2 &=& -(0.1558 \pm 0.0098) \nonumber
\end{eqnarray}
and has a colour zero-point 0.03--0.04 mag bluer than the corresponding {\tt model} magnitudes CMR. This again is due to colour gradients, which cause $g$-band effective radii to be systematically larger than those fitted in the $r$-band. However, the CMR slope is not significantly different, which suggests the mean ratio of (observer-frame) $g$ and $r$ radii does not strongly depend on luminosity. The greater blueward evolution of the de Vaucouleurs CMR may mean that this ratio increases with redshift (from gradient evolution, or from steeper gradients at shorter rest-frame wavelengths, e.g. $u-b$). 
 \subsection{CMR from other $k$-corrections} 

\begin{figure}
 \includegraphics[width=0.7\hsize,angle=-90]{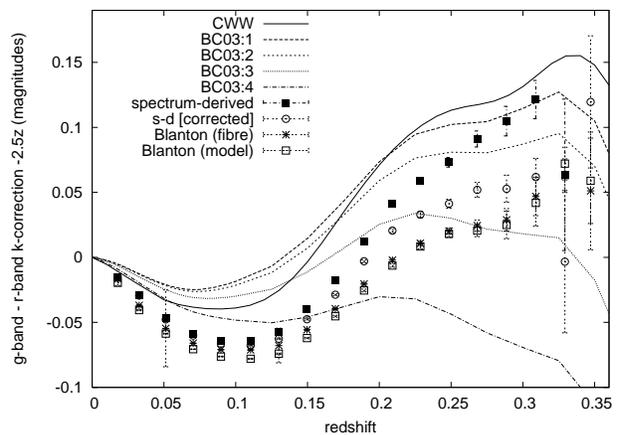} 
\caption{The difference $k_g-k_r$, with a linear $-2.5z$ trend subtracted, for E/S0 galaxies in $\Delta(z)=0.025$ bins, from the spectra (with and without the signal/noise correction), from Blanton \& Roweis (2007) for {\tt model} and {\tt fiber} magnitudes), from the 4 BC03 models, and from CWW.} 
 \label{kgkr}
\end{figure} 

In this subsection, we consider the k-correction computed following Blanton \& Roweis (2007), a program which fits the broad-band  ($ugriz$) magnitudes of each individual galaxy with one of a set of BC03 model template spectra, from which the $g$ and $r$-band k-corrections are then measured. We also compare with two simpler models in which the same k-corrections are assumed for all the E/S0s:  firstly, our evolving BC03 model 2 (age 12 Gyr $\tau=1$ Gyr) and secondly, the non-evolving for ellipticals from Coleman, Wu and Weedman (1980, hereafter CWW).
Figure~\ref{kgkr} shows the different $k_g-k_r$ against redshift (Table~1\&2 list the different $k$-corrections used in the Figure).

 Firstly, note that k-corrections computed from {\tt model} and {\tt fiber} magnitudes differ (primarily at $z < 0.15$) by less than 0.01 mag. At $z<0.15$ the spectra and Blanton-Roweis $k_g-k_r$ are all below our BC Model 4, but the uncorrected $k_g-k_r$ crosses all the models to reach Model~1 at $z>0.3$. The corrected $k_g-k_r$ is lower at high redshift, and the Blanton-Roweis $k_g-k_r$ are lower still, and give the least evolution. The CWW $k_g-k_r$ is always more positive than that from the spectra, and so gives an overestimate of $\Delta(g-r)$ and too-blue rest-frame colours. 

\begin{figure}
\includegraphics[width=0.7\hsize,angle=-90]{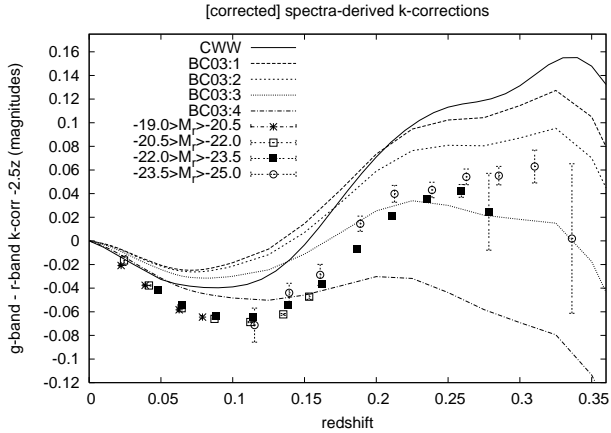} 
\caption{Same as previous figure, but now 
% The mean spectra-derived k-correction for $g-r$ (with the signal/noise correction) with a linear trend $-2.5z$ subtracted, 
showing the corrected spectra-derived k-correction
for E/S0 galaxies divided into 4 intervals of absolute magnitude $M_r$.} 
 \label{kgkr-fixedM}
\end{figure} 

Neither the corrected or uncorrected $k_g-k_r$ follow any of the BC03 models over the whole redshift range. This is to some extent because the mean luminosity increases with redshift. To examine the effect of this, Figure~\ref{kgkr-fixedM} shows mean $k_g-k_g$ with the galaxies divided into four intervals of $M_r$. At a given redshift the more luminous galaxies do tend to have a slightly more positive $k_g-k_r$.  However, the galaxy $k_g-k_r$ within each of the four luminosity intervals still do not closely follow any of the BC03 models, tending to be lower at $0.05<z<0.2$, indicating systematic differences at some wavelengths between the spectra and this set of models.

\begin{figure}
 \includegraphics[width=0.7\hsize,angle=-90]{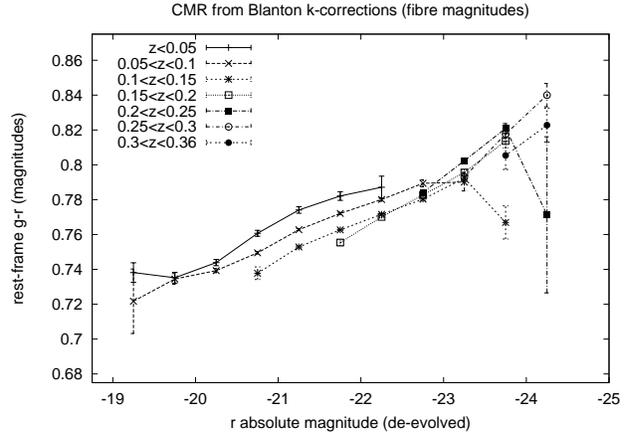} 
 \includegraphics[width=0.7\hsize,angle=-90]{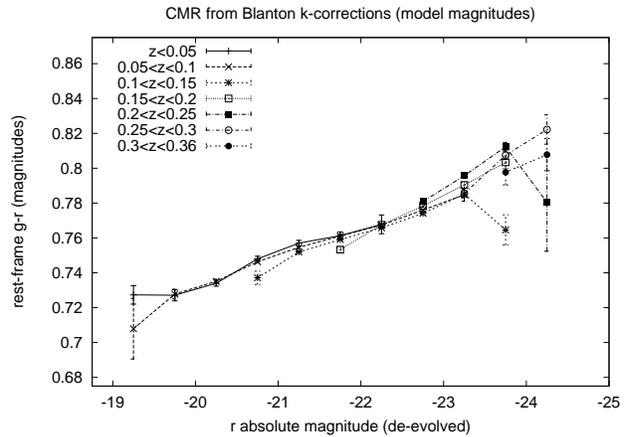} 
 \caption{Colour-magnitude relation using $k$-corrections from Blanton \& Roweis (2007) fitted to $ugriz$ {\tt fiber} (top) and {\tt model} magnitudes (bottom).}
 \label{cmr-br}
 \end{figure}

We compared the spectroscopically-based CMRs with the CMRs from photometric colours $k$-corrected using only model-based or template spectra. The top panel in Figure~\ref{cmr-br} shows the effect of fitting Blanton \& Roweis (2007) $k$-corrections to {\tt fiber} magnitudes, and using the observed {\tt fiber} $g-r$.  The CMR is best-fit with
\begin{eqnarray}
 a_0 &=& 0.2852\pm 0.0080 \nonumber\\
 a_1 &=& -(0.0227\pm 0.00039)\nonumber \\
 a_2 &=&  -(0.10541 \pm 0.0064)\nonumber;
\end{eqnarray}
The slope is very similar to that in the bottom panel of Figure~\ref{cmr-fiber}, but with these rather than the spectra-based $k$-corrections, the evolution is $0.09z$ weaker.  A similar analysis of the {\tt model} colour (Blanton \& Roweis $k$-corrections from the {\tt model} $ugriz$ magnitudes) is shown in the bottom panel of Figure~\ref{cmr-br}.  Again the model-magnitude colours give a CMR with a shallower slope,
\begin{eqnarray}
 a_0 &=& 0.3781\pm 0.0075 \nonumber\\
 a_1 &=&  - (0.01761\pm 0.00037)\nonumber \\
 a_2 &=&  -(0.0029 \pm 0.0060),\nonumber
\end{eqnarray}
but with these k-corrections the evolution is reduced to zero or even inverted!
This may be an artifact of the fact that the BC03 models used by Blanton \& Roweis assume solar abundance ratios, and this is unrealistic.  

 \begin{figure}
 \includegraphics[width=0.7\hsize,angle=-90]{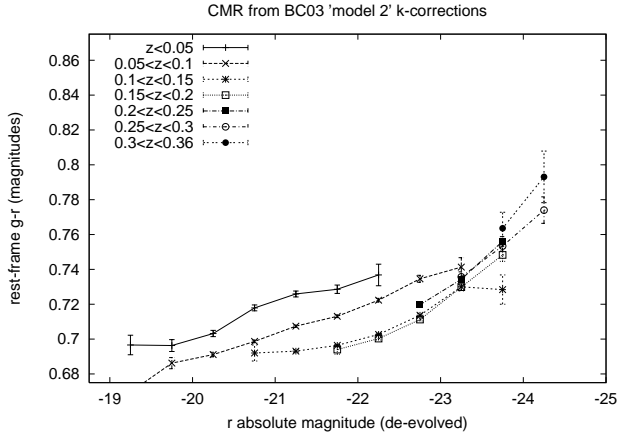}
\caption{Colour-magnitude relation for E/S0s calculated using {\tt model} magnitudes and $k$-corrections from our BC03:2 model (age 12~Gyr and $\tau=1$ Gyr).}
 \label{cmr-bc03}
\end{figure}

 \begin{figure}
 \includegraphics[width=0.7\hsize,angle=-90]{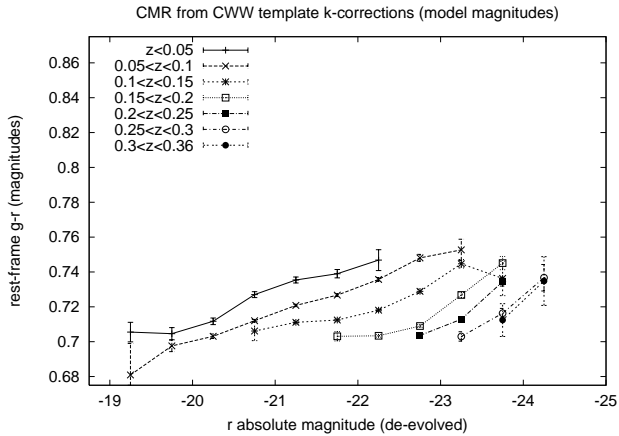}
\caption{Colour-magnitude relation for E/S0s calculated using {\tt model} magnitudes and $k$-corrections the CWW non-evolving elliptical template spectrum.}
 \label{cmr-cww}
\end{figure}

Figure~\ref{cmr-bc03} shows the CMR from {\tt model} colours and $k$-corrections from BC03 Model 2 k-corrections.  It is well fit by equation~(\ref{cmrfit}) with 
\begin{eqnarray}
 a_0 &=& 0.3241\pm 0.0094 \nonumber\\
 a_1 &=& -(0.01846\pm 0.00046)\nonumber \\
 a_2 &=& -(0.1725 \pm 0.0076)\nonumber .
\end{eqnarray}
Although this evolution rate seems consistent with the $\Delta(g-r)=-0.17z$  directly predicted by the model, there is a discrepancy in that all the CMR evolution is at $z<0.15$. Indeed, Figures~\ref{bc03colours} and~\ref{bc03kcorr} showed that no single BC03 model fits the colours and k-corrections of both the low and high redshift E/S0s.  

The result of using the CWW elliptical galaxy template to compute the $k$-correction is shown in Figure~\ref{cmr-cww}.  This CMR has 
\begin{eqnarray}
 a_0 &=& 0.3312\pm 0.0094\nonumber\\
 a_1 &=& -(0.01955\pm 0.00046)\nonumber \\
 a_2 &=& -(0.3530 \pm 0.0076) \nonumber ; 
\end{eqnarray}
this produces rather more evolution than expected from the BC03 models.  

It is evident the CMR and its evolution, as derived from the photometry, are extremely sensitive to the adopted $k$-correction.  For example, both the BC03 and CWW $k$-corrections give bluer rest-frame colours than the either the spectra or Blanton-Roweis models. 
 
\section{The colour-$\sigma$ relation (C$\sigma$R)} 

\begin{figure}
 \includegraphics[width=0.7\hsize,angle=-90]{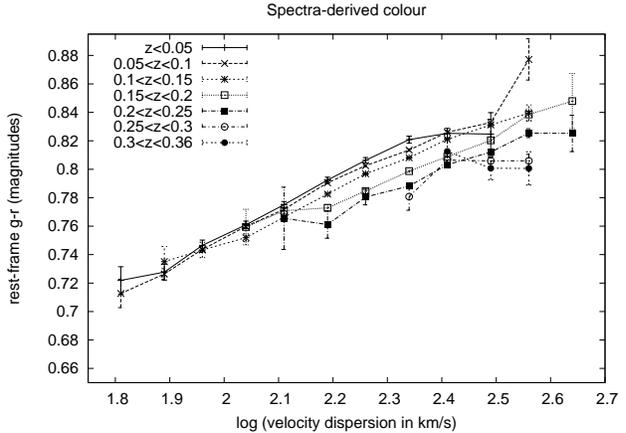} 
 \includegraphics[width=0.7\hsize,angle=-90]{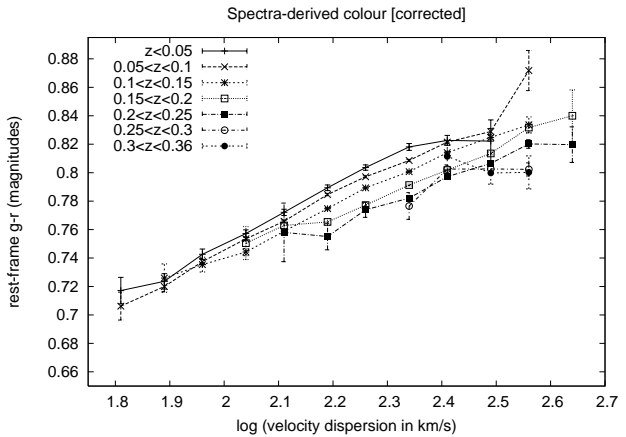} 
 \caption{C$\sigma$R using rest-frame $g-r$ derived from the spectra, before and after the signal/noise correction (top and bottom).  Points corresponding to fewer than 8 galaxies are not plotted.} 
 \label{cvr-spec}
\end{figure} 

\begin{figure}
 \includegraphics[width=0.7\hsize,angle=-90]{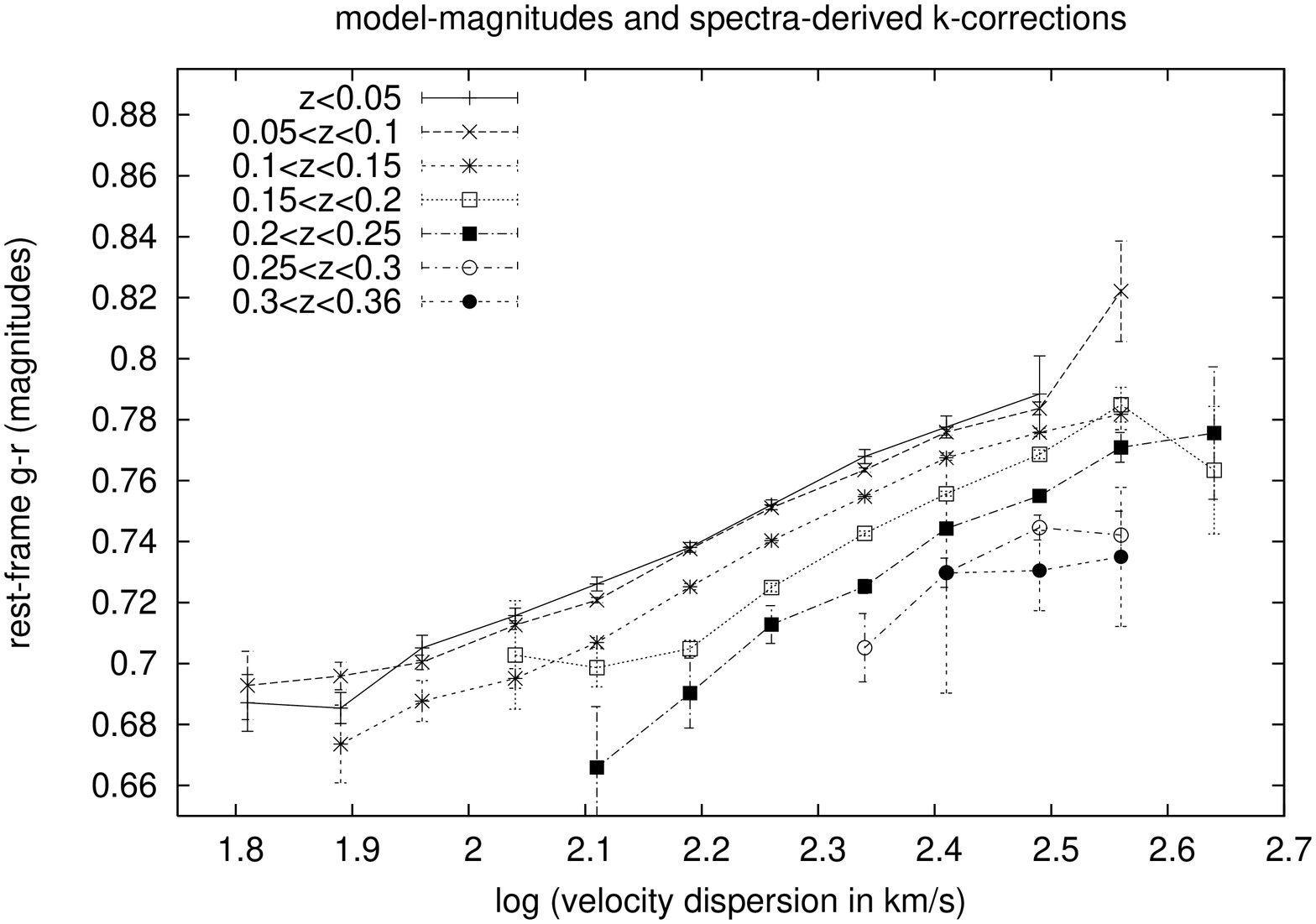} 
 \includegraphics[width=0.7\hsize,angle=-90]{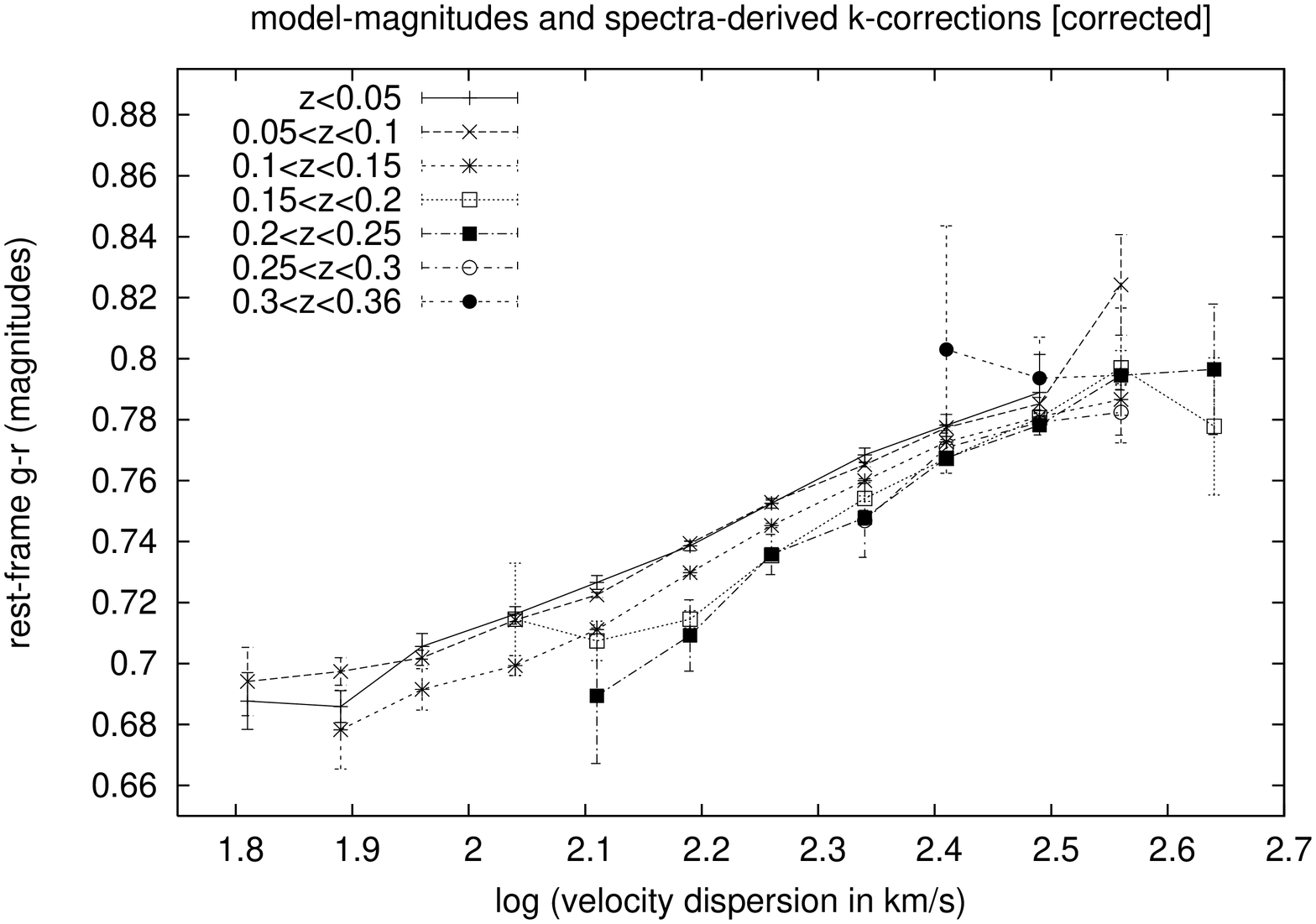} 
 \caption{Same as previous figure, but now using {\tt model} $g-r$ colour and $k$-corrections from the spectra, before and after the signal/noise correction.}
 \label{cvr-model}
\end{figure}

\begin{figure}
 \includegraphics[width=0.7\hsize,angle=-90]{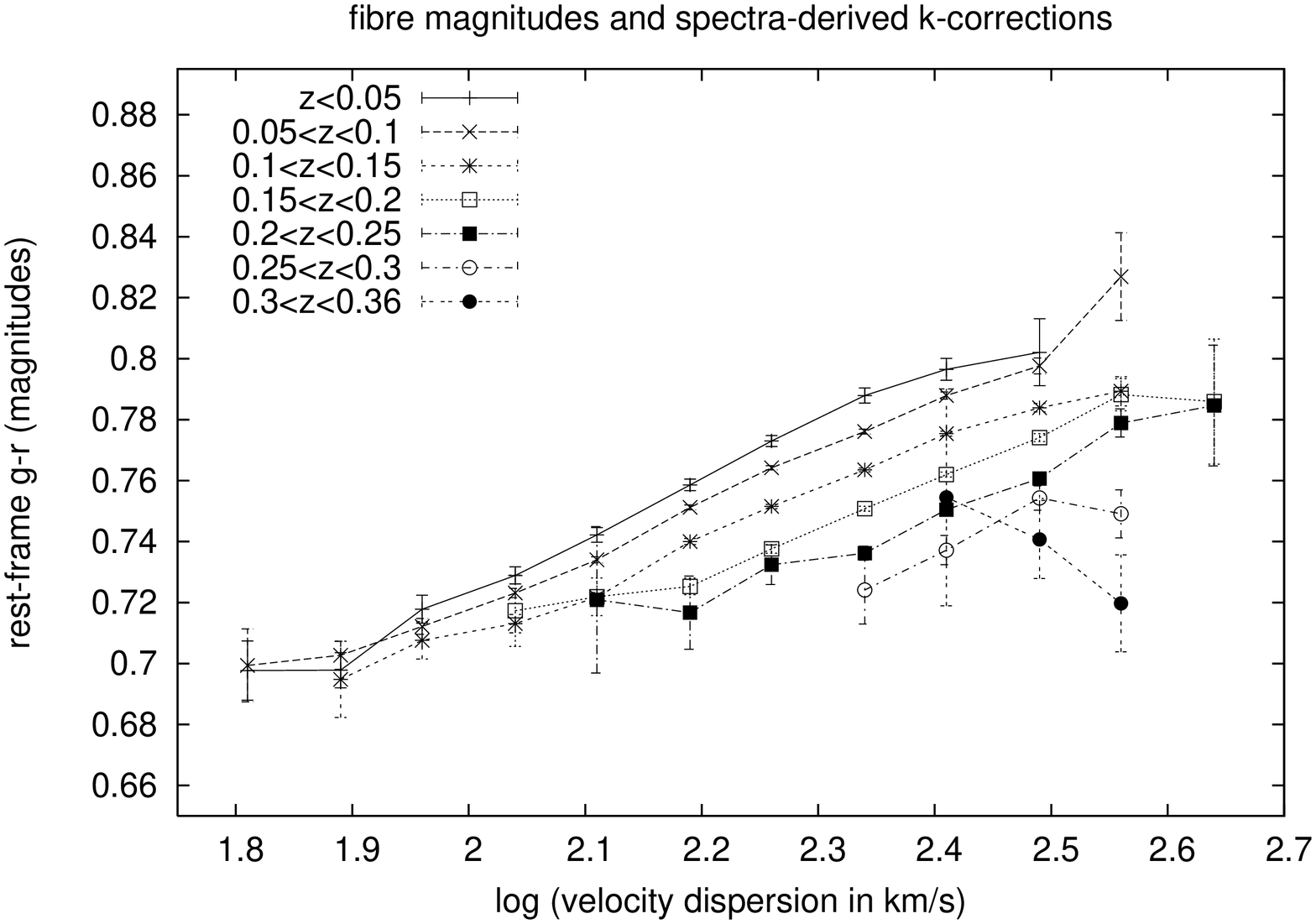} 
 \includegraphics[width=0.7\hsize,angle=-90]{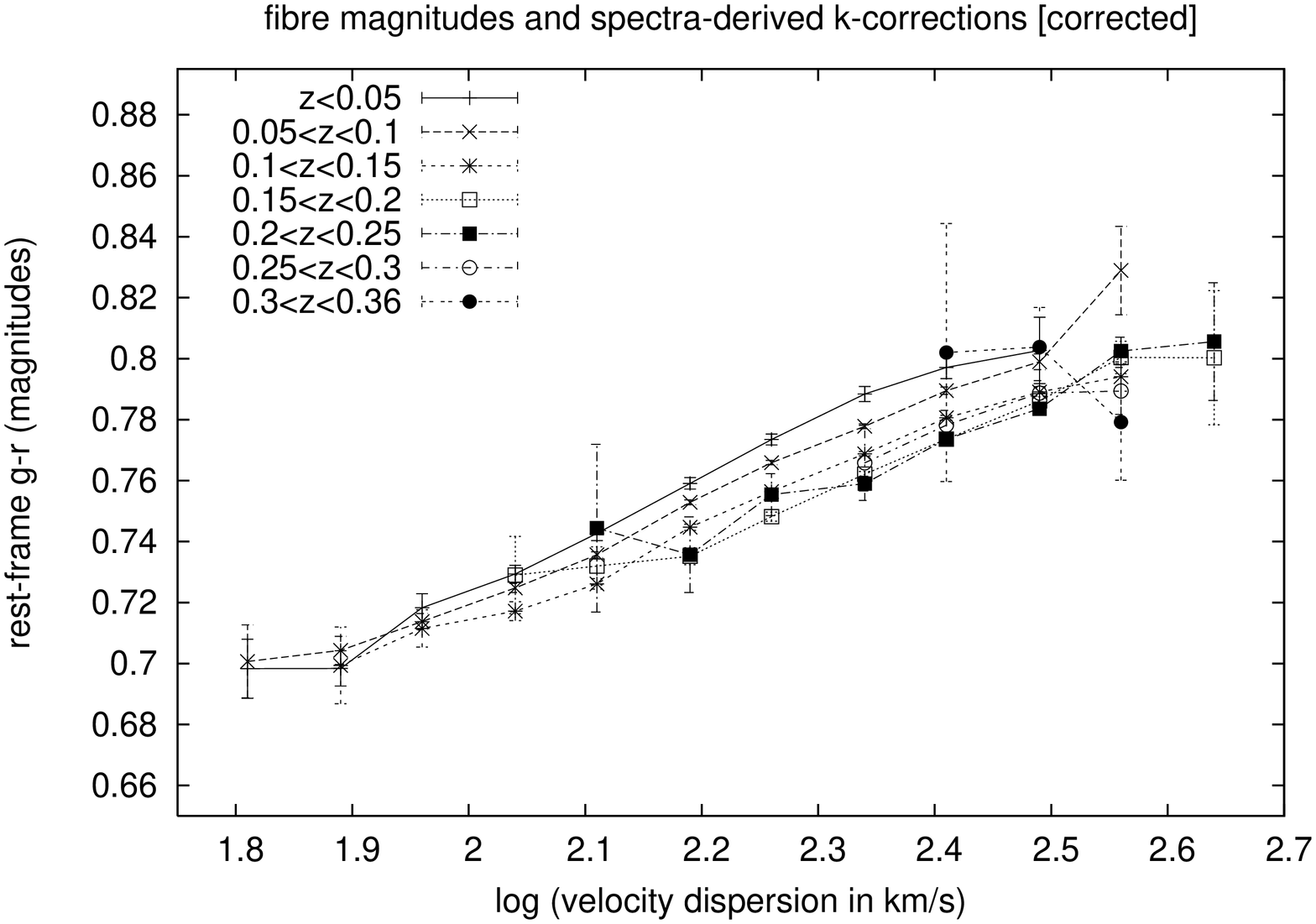} 
 \caption{Same as previous figure, but now using {\tt fiber} colours and $k$-corrections from the spectra, before and after the signal/noise correction.}
 \label{cvr-fiber}
\end{figure}

The CMR is due to more fundamental correlations between colour and velocity dispersion $\sigma$, and $\sigma$ and luminosity (Bernardi et al. 2005).  The previous section showed that colour gradients affect the zero-point and slope of the CMR.  We now study how the colour-$\sigma$ relation (hereafter C$\sigma$R) depends on the aperture within which the colour is defined. 

Figure~\ref{cvr-spec} shows the C$\sigma$R for our sample for rest-frame $g-r$ from the spectra, before (top) and after (bottom) correcting for signal/noise.  Fitting the galaxies by equation~(8), but with $M_r$ replaced by $\log\sigma$ yields 
\begin{eqnarray}
 a_0 &=& 0.3934\pm 0.0041 \quad{\rm or}\quad 0.3804\pm 0.0040\nonumber\\ 
 a_1 &=& 0.18685\pm 0.00195 \quad{\rm or}\quad 0.19061\pm 0.00189\nonumber \\
 a_2 &=& - (0.1860\pm 0.0049) \quad{\rm or}\quad -(0.2032\pm 0.0048)\nonumber
\end{eqnarray}
with rms residuals of 0.0521 magnitudes. 

\begin{figure} 
\includegraphics[width=70mm,angle=-90]{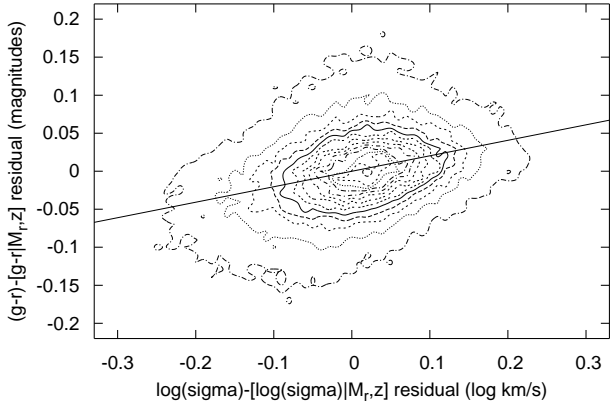} 
\caption{The residual $(g-r)_{rf}-\langle (g-r)_{rf}|M_r,z \rangle $
 plotted against the 
residual ${\rm log}\sigma-\langle {\rm log}\sigma|M_r,z\rangle$, 
as a contour plot with the best-fit linear regression.} 
\label{resid1}
\end{figure}

\begin{figure} 
\includegraphics[width=70mm,angle=-90]{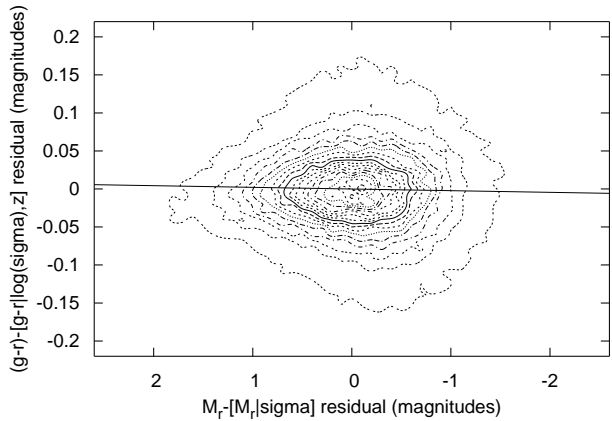} 
\caption{The residual 
$(g-r)_{rf}-\langle (g-r)_{rf}|\sigma,z \rangle$ plotted against the 
residual $M_r-\langle M_r|\sigma,z\rangle$.}
\label{resid2}
\end{figure}

Note that colour evolution is strongly detected, but it is $\sim 0.06$--$0.07z$ less than in the corresponding CMR.  
Using {\tt model} $g-r$ and the (uncorrected or corrected) spectra-derived k-corrections yields 
\begin{eqnarray}
 a_0 &=& 0.3317\pm 0.0043  \quad{\rm or}\quad 0.3285\pm 0.0044\nonumber\\ 
 a_1 &=& 0.19416\pm 0.00206 \quad{\rm or}\quad 0.19139\pm 0.00209\nonumber \\
 a_2 &=&  - (0.2607\pm 0.0052) \quad{\rm or}\quad - (0.1289\pm 0.0053)\nonumber
\end{eqnarray}
with rms residuals of 0.0551 magnitudes (see Figure~\ref{cvr-model}).
This C$\sigma$R relation is bluer than that based on the spectra, but the slope is almost unchanged.  However, it has $0.13z$ less evolution.  
Finally, using {\tt fiber} $g-r$ (Figure~\ref{cvr-fiber}) instead gives 
\begin{eqnarray}
 a_0 &=& 0.3740\pm 0.0045 \quad{\rm or}\quad 0.3709\pm 0.0046 \nonumber\\ 
 a_1 &=& 0.18186\pm 0.00212 \quad{\rm or}\quad 0.17902\pm 0.00215\nonumber \\
 a_2 &=&  -(0.2859\pm 0.0053) \quad{\rm or}\quad - (0.1540\pm 0.0040)\nonumber
\end{eqnarray}
 Following Bernardi et al. (2005), if we plot the residual of $g-r$ colours ({\tt model} magnitude and signal/noise corrected) to the CMR, $(g-r)_{rf}-\langle (g-r)_{rf}|M_r,
z\rangle$, against the residual of {\rm log}~$\sigma$ to the best-fit 
$\sigma-M_r$ relation, we see the two are correlated (Figure ~\ref{resid1}) . The correlation coefficient is 0.2994 with a slope $0.2033\pm 0.0024$, an intercept $0.0000\pm 0.0002$, and a 
standard deviation 0.0559 mag. However, if we plot the residual of colours to the C$\sigma$R, $(g-r)_{rf}-\langle (g-r)_{rf}|\sigma,z \rangle$ against the residual of $M_r$ to the best-fit $M_r-\sigma$ relation (Figure ~\ref{resid2})), we find a much lower  correlation coefficient 0.0231 and slope  $0.0022\pm 0.0004$
(with an intercept again
$0.0000\pm 0.0002$ and standard deviation 0.0560 magnitudes).

Again we find that $g-r$ colour in E/S0s are more fundamentally dependent on velocity dispersion than on $M_r$, and furthermore, the absence of an $M_r$ correlation in the second case implies that the C$\sigma$R is the same for all bins in magnitude at a fixed velocity dispersion. Hence our determination of the C$\sigma$R in this Section should not be significantly biased by our use of a flux-limited sample. 


In both the CMR and C$\sigma$R, rest-frame colours from the spectra show stronger $g-r$ evolution (0.20--$0.27z$) than model-magnitude colours with the `corrected' k-correction (0.07--$0.13z$).  This is an aperture affect, due to the colour gradients which also shift the zero-point of both the CMR and C$\sigma$R bluewards when {\tt model} colours are used.  However, it is notable that, while the CMR based on {\tt model} colours is $30\%$ shallower than the CMR based on colours from within 3 arcsec, this does not happen for the C$\sigma$R. This suggests that the strength of the colour gradient is positively correlated with luminosity, but not with velocity dispersion.

\begin{figure} 
 \includegraphics[width=70mm,angle=-90]{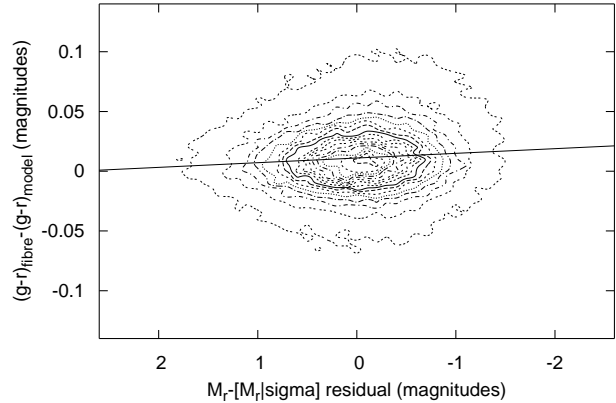} 
 \caption{The difference in rest-frame $g-r$ colours derived using fiber magnitudes and model magnitudes (with k-corrections from the spectra), plotted against the residual $M_r-\langle M_r|\sigma,z\rangle$.}
 \label{grad-resid}
\end{figure}

We study this in Figure~\ref{grad-resid}, which shows the difference between the rest-frame $g-r$ from {\tt fiber} and {\tt model} magnitudes against the residual of $M_r$ to the best-fit $M_r-\sigma$ relation. We find there is a weak but significant correlation, with correlation coefficient 0.069, slope $-0.0039\pm 0.0002$, intercept $0.0111\pm 0.0001$,  and a standard deviation 0.0327 magnitudes. The slope $d(g-r)/d M_r\simeq -0.0039$, can account for the greater part of the additional steepness of the fiber and spectra magnitude CMRs relative to the model magnitude CMR. 
\begin{figure} 
 \includegraphics[width=70mm,angle=-90]{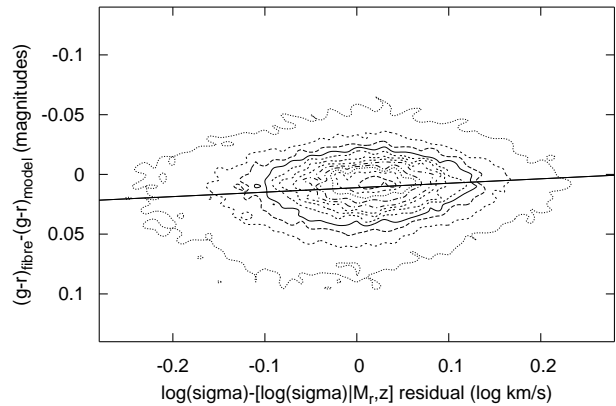} 
 \caption{The difference in rest-frame $g-r$ colours derived using fiber and model magnitudes (with k-corrections from the spectra), plotted against the residual ${\rm log}~\sigma-\langle {\rm log}~\sigma|M_r,z\rangle$.}
 \label{grad-resid2}
\end{figure}
Similarly, the difference between the rest-frame $g-r$ from {\tt fiber} and {\tt} model magnitudes can be plotted against the residual of ${\rm log}~\sigma$ to the best-fit $\sigma-M_r$ relation (Figure~\ref{grad-resid2}). We note that this residual is positively correlated to galaxy age (Bernardi et al. 2005; Gallazzi et al. 2006). Again we find a correlation, with coefficient 0.098, slope $-0.0372\pm 0.0014$, intercept $0.0111\pm 0.0001$,  and a standard deviation 0.0326 magnitudes.
These results indicate a dependence of colour gradient in E/S0s on the galaxy properties, such as age. This is the subject of a work in progress.  
 
\section{Summary and Discussion} 
\medskip
\noindent 1. From the SDSS data release 6, with 367471 galaxy spectra, we selected 70378 E/S0 (early-type) galaxies at $0<z<0.36$, using a combination of both morphological (pure de Vaucouleurs profile) and spectral (negative {\tt eclass} from principal component analysis) criteria. We measured observer-frame and rest-frame $g$ and $r$-band magnitudes for these galaxies by integrating over the SDSS spectra (corrected for atmospheric and Galactic absorption) weighting by the appropriate filter functions.  The spectra magnitudes of these galaxies were then compared with photometric (3 arcsec aperture) {\tt fiber} magnitudes from the SDSS imaging data. There was a large scatter ($\sim 0.1$ mag) between the two but on average, the spectra gave systematically redder $g-r$ colours than the {\tt fiber} magnitudes, the difference being 0.03--0.04 magnitudes at lower redshifts.

Some, but not all, of this discrepancy is due to seeing:  the flux in the SDSS spectra is reduced relative to the imaging photometry by about 0.021 mag per arcsec of seeing in $g$ and 0.012 mag in $r$. As the median seeing is 1.8 arcsec this may cause $\Delta (g-r)\sim 0.01$ mag reddening of the spectra.  There may be a additional small systematic differences in the colour calibrations, consistent with the quoted uncertainties of up to 0.02 mag (Adelman-Mccarthy et al. 2008).

\medskip
\noindent 2.  In the $g$ band, the mean difference between $g_{spec}$ and $g_{fib}$ decreases significantly ($\sim 0.05$ mag) with increasing redshift, although the scatter increases.  A similar trend is {\em not} seen in the $r$ band, potentially compromising estimates of how the colours of these objects evolve.  The redshift dependence of $g_{spec}-g_{fib}$ with redshift is much greater for galaxies with the most negative values of  the {\tt eclass} spectral classification parameter.  At $z\sim 0.3$, such objects have low flux and high noise at the blue end of the $g$-band, and $g_{spec}-g_{fib}$ increases at low signal/noise.  We calibrated this dependence on {\tt eclass} and signal/noise, and used it to `correct' spectra-derived magnitudes for these effects.  Understanding the underlying cause for these trends requires further work. For example, a comparison of SDSS spectra with best-fitting BC03-type models, stacking large numbers together to increase signal/noise, might help to reveal exactly where in the spectra the problem is occurring.

\medskip
\noindent 3. The differences between observer-frame and rest-frame magnitudes as measured from the spectra provide estimates of $k$-corrections in $g$ and $r$. %This correction reduced the mean $g$-band correction significantly at higher redshift, by 0.05 mag at $z=0.3$ but it had a minimal effect of $<0.01$ magnitudes on the mean k-correction in the $r$-band, where the signal/noise was generally much higher.
The k-corrections from the SDSS spectra of E/S0s were similar to those from BC03 models with ages $\sim 12$ Gyr, and star-formation exponentially declining with $\tau=1$ Gyr for the most luminous galaxies and $\tau=2$ Gyr for the least luminous. 
%Without the correction the most luminous galaxies may be closer to a burst model. 
These SF histories and their moderate luminosity dependence agree with Jimenez et al. (2007). 

\medskip
\noindent 4. Rest-frame colours $(g-r)_{rf}$ were measured directly from the spectra, and also calculated by subtracting $k$-corrections from the SDSS {\tt fiber} or {\tt model} magnitudes.  The first two colour estimates are based on fixed angular apertures (about 3 arcsec); the scale associated with the latter is related to the half-light radius, so it varies from object to object.  
Absolute magnitudes $M_r$ were calculated from the $k$-corrected {\tt model} magnitudes, and were then corrected for evolution.  
%Using a $V/V_{max}$ method, we estimated the evolution of the E/S0 characteristic absolute magnitude as $\Delta(M^*_r)\simeq -0.86z$.
The rest-frame colour-magnitude relation (CMR) and colour-$\sigma$ (C$\sigma$R) relations were measured in a series of redshift intervals.  Both are well-approximated by power-laws that tend to shift gradually bluewards with increasing redshift.  The zero-point, slope and evolution of the CMR associated with fixed angular apertures (spectral and {\tt fiber} colours) differ from that found for {\tt model} magnitudes. We found the slope  30--40\% steeper, and note Mei et al. (2006) and Scodeggio (2001) found similar differences in the CMRs from fixed aperture or from half-light colours. These differences are a consequence of radial colour gradients.  Gradients complicate interpretation of evolution of the CMR from fixed apertures -- evolution of the CMR based on {\tt model} magnitudes is easier to interpret.  

\medskip
\noindent 5.  Whereas colour gradients affect the zero-point of the C$\sigma$R, the slope is much less affected.  This suggests that gradients are not positively correlated with $\sigma$ -- but we find indications that they are correlated  with residuals in the high luminosity and low $\sigma$ direction from the $\langle\sigma|M_r,z\rangle$ relation.

\medskip
\noindent 6. The $k$-corrections obtained from fitting BC03 template spectra to the observed $ugriz$ fluxes (following Blanton \& Roweis 2007) depend only slightly on whether one uses {\tt fiber} or {\tt model} magnitudes ($< 0.01$ mag at $z < 0.15$).

\medskip
\noindent 7.  We also calculated the CMR using k-corrrections derived from model or template spectra, rather than the SDSS spectra.  The estimate of rest-frame colour evolution is very sensitive to the assumed $k$-correction, and these models gave very different results.  The BC03 Model 2 gave a reasonable $(g-r)_{rf}$evolution rate $-0.1662z$, but at an uneven rate with all the evolution at $z<0.15$, apparently because a model with a single SF history could not simultaneously fit the more and less luminous galaxies. The CWW $k$-correction, which is based on a non-evolving spectrum, gave a CMR with strong evolution of $-0.3589z$, so this did not give self-consistent results. The $k$-corrections from Blanton \& Roweis (2007) are smaller than those based on the SDSS spectra, and so the CMRs show less evolution -- the CMR based on {\tt model} magnitudes showed no evolution.  This may be an artifact of the fact that the BC03 models assume solar abundance ratios, and this is unrealistic.  

 \section*{Acknowledgements} We thank David Schlegel and R. Sheth for interesting discussions about our work, which was supported in part by NASA grant LTSA-NNG06GC19G.

Funding for the Sloan Digital Sky Survey (SDSS) has been provided by the Alfred P. Sloan Foundation, the Participating Institutions, the National Aeronautics and Space Administration, the National Science Foundation, the U.S. Department of Energy, the Japanese Monbukagakusho, and the Max Planck Society. The SDSS Web site is http://www.sdss.org/. The SDSS is managed by the Astrophysical Research Consortium (ARC) for the Participating Institutions. The Participating Institutions are The University of Chicago, Fermilab, the Institute for Advanced Study, the Japan Participation Group, The Johns Hopkins University, the Korean Scientist Group, Los Alamos National Laboratory, the Max-Planck-Institute for Astronomy (MPIA), the Max-Planck-Institute for Astrophysics (MPA), New Mexico State University, University of Pittsburgh, University of Portsmouth, Princeton University, the United States Naval Observatory, and the University of Washington.

\section*{References} 
\vskip0.15cm \noindent Adelman-McCarthy et al., 2006, ApJS, 162, 38.

\vskip0.15cm \noindent Adelman-McCarthy et al. 2008, ApJS, 175, 297.

\vskip0.15cm \noindent Baum W.A., 1959, Pub. Ast. Soc. Pacif., 71, 106.

\vskip0.15cm \noindent Bernardi M., et al. 2003a, AJ, 125, 1817.

\vskip0.15cm \noindent Bernardi M., et al. 2003b, AJ, 125, 1849.

\vskip0.15cm \noindent Bernardi M., et al. 2003c, AJ, 125, 1882.

\vskip0.15cm \noindent Bernardi M., Sheth R.K., Nichol R.C., Schneider 
D.P., Brinkmann J., 2005, AJ, 129, 61.

\vskip0.15cm \noindent Bernardi M., Nichol R.C., Sheth R.K., Miller 
C.J., Brinkmann J., 2006, AJ, 131, 1288.

\vskip0.15cm \noindent Bernardi M, Hyde J.B., Sheth R.K., Miller C.J., 
Nichol R.C., 2007, AJ, 133, 1741.

\vskip0.15cm \noindent Bernardi M, Hyde J.B., Fritz A., Sheth R.K., Gebhardt K., Nichol R.C., 2008, MNRAS, 391, 1191.

\vskip0.15cm \noindent Blakeslee J.P., et al., 2003, ApJ, 596, L143.

\vskip0.15cm \noindent Blanton M.R. \& Roweis S., 2007, AJ, 133, 734.

\vskip0.15cm \noindent Bower R.G., Lucey J.R., Ellis R.S., 1992, MNRAS, 
254, 601.

\vskip0.15cm \noindent Bruzual A.G. and Charlot S., 2003, MNRAS, 344, 
1000.

\vskip0.15cm \noindent Chabrier G., 2003, PASP, 115, 763.

\vskip0.15cm \noindent Coleman G.D., Wu C. and Weedman D.W., 1980, ApJS, 
94, 63.

\vskip0.15cm \noindent Gallazzi A., Charlot S., Brinchmann J., White 
S.D.M., 2006, MNRAS, 370, 1106.

\vskip0.15cm \noindent Hogg D., Baldry I., Blanton M., Eisenstein D., 2002, astro-ph/0210394

\vskip0.15cm \noindent Holden B.P., Stanford S.A., Eisenhardt P., Dickinson M., 2004, AJ, 127, 2484.

\vskip0.15cm \noindent Hyde J. and Bernardi M., 2009, MNRAS, in press.
astro-ph/0810.4922
 
\vskip0.15cm \noindent Jimenez R., Mariangela B., Haiman Z., Panter B., 
Heavens A.F., 2007, ApJ, 669, 947.

\vskip0.15cm \noindent J{\/o}rgensen I., Franx M., Kjaergaard P., 1995, MNRAS, 276, 1341.

\vskip0.15cm \noindent Kodama T. and Arimoto N., 1997, A\&A, 320,41.

\vskip0.15cm \noindent Kroupa P., 2001, MNRAS, 322, 231.

\vskip0.15cm \noindent Labb\'e I. et al., 2007, ApJ 665, 944.

\vskip0.15cm \noindent Lauer T.R. et al., 2007, ApJ, 662, 808.

\vskip0.15cm \noindent Mei, Simona, et al., 2006, ApJ, 644,759.

\vskip0.15cm \noindent Mei, Simona, et al., 2009, ApJ, 690,42.

\vskip0.15cm \noindent Petrosian V., 1976, ApJ, 201, L1.

\vskip0.15cm \noindent Salviander S., Shields G.A., Gebhardt K., Bernardi M., Hyde J.B., 2008, ApJ, 687, 828.

\vskip0.15cm \noindent Scodeggio M., 2001, AJ, 121, 2413.

\vskip0.15cm \noindent Stoughton C. et al. 2002, AJ, 123, 485.

\vskip0.15cm \noindent Taylor E.N., et al. 2008, ApJ, in press (astro-ph/0810.3459).

\vskip0.15cm \noindent de Vaucouleurs, G. et al. 1948, Ann d'Astrophys., 
11, 247.

\vskip0.15cm \noindent Vazdekis A., Kuntschner H., Davies R.L., Arimoto 
N, Nakamura O., Peletier R., 2001, ApJ, 551, L127.

\vskip0.15cm \noindent van Dokkum P.G., Franx M., Fabricant D., Illingworth G.D., Kelson D.D., 2000, ApJ, 541, 95.

\vskip0.15cm \noindent Visvanathan N., Sandage A., 1977, ApJ, 216, 214.

\vskip0.15cm \noindent Yip C.W., et al. 2004, AJ, 128, 585.

\onecolumn
\begin{table}
\caption{$g$-band k-corrections for E/S0 galaxies (averaged in redshift intervals), derived from the SDSS spectra (with and without the signal/noise correction, the Blanton models (based on fibre or model magnitudes) and the CWW non-evolving elliptical galaxy template.} 
\begin{tabular}{lccccc}
\hline
Redshift & \multispan{2} Spectra-derived & \multispan{2} Blanton~model & CWW elliptical \\
\smallskip
 range   & uncorrected & corrected & fibre mag. & model mag. &  template \\
0.0--0.02  & 0.0467 & 0.0465 & 0.0453 & 0.0421 & 0.0544 \\
0.02--0.04 & 0.0877 & 0.0871 & 0.8216 & 0.0775 & 0.0992 \\ 
0.04--0.06 & 0.1398 & 0.1385 & 0.1135 & 0.1274 & 0.1557 \\
0.06--0.08 & 0.2006 & 0.1983 & 0.1944 & 0.1859 & 0.2207 \\
0.08--0.10 & 0.2690 & 0.2645 & 0.2610 & 0.2520 & 0.2953 \\
0.10--0.12 & 0.3526 & 0.3474 & 0.3410 & 0.3308 & 0.3851 \\
0.12--0.14 & 0.4389 & 0.4317 & 0.4128 & 0.4129 & 0.4780 \\
0.14--0.16 & 0.5329 & 0.5230 & 0.5087 & 0.5001 & 0.5756 \\
0.16--0.18 & 0.6324 & 0.6193 & 0.6021 & 0.5944 & 0.6784 \\
0.18--0.20 & 0.7406 & 0.7236 & 0.6993 & 0.6932 & 0.7853 \\
0.20--0.22 & 0.8500 & 0.8274 & 0.7965 & 0.7908 & 0.8887 \\
0.22--0.24 & 0.9470 & 0.9191 & 0.8870 & 0.8828 & 0.9854 \\
0.24--0.26 & 1.0454 & 1.0115 & 0.9760 & 0.9724 & 1.0787 \\
0.26--0.28 & 1.1480 & 1.1071 & 1.0638 & 1.0584 & 1.1688 \\ 
0.28--0.30 & 1.2641 & 1.2104 & 1.1583 & 1.1537 & 1.2653 \\
0.30--0.32 & 1.3614 & 1.3009 & 1.2564 & 1.2461 & 1.3682 \\
0.32--0.34 & 1.4141 & 1.3456 & 1.3721 & 1.3737 & 1.4717 \\
0.34--0.36 & 1.6124 & 1.5301 & 1.4303 & 1.4229 & 1.5524 \\
\hline
\end{tabular}
\end{table}
\medskip
\begin{table}
\caption{As above, for the $r$-band k-corrections}
\begin{tabular}{lccccc}
\hline
Redshift & \multispan{2} Spectra-derived & \multispan{2} Blanton~model & CWW elliptical \\
\smallskip
 range   & uncorrected & corrected & fibre mag. & model mag. &  template \\
0.0--0.02  & 0.0174 & 0.0173 & 0.0170 & 0.0170 & 0.0206 \\
0.02--0.04 & 0.0348 & 0.0346 & 0.0372 & 0.0361 & 0.0384 \\ 
0.04--0.06 & 0.0585 & 0.0580 & 0.0399 & 0.0578 & 0.0600 \\
0.06--0.08 & 0.0834 & 0.0825 & 0.0845 & 0.0805 & 0.0828 \\
0.08--0.10 & 0.1090 & 0.1077 & 0.1078 & 0.1041 & 0.1105 \\
0.10--0.12 & 0.1408 & 0.1392 & 0.1362 & 0.1330 & 0.1454 \\
0.12--0.14 & 0.1711 & 0.1693 & 0.1555 & 0.1619 & 0.1778 \\
0.14--0.16 & 0.1985 & 0.1965 & 0.5087 & 0.5001 & 0.5756 \\
0.16--0.18 & 0.2266 & 0.2246 & 0.2181 & 0.2163 & 0.2292 \\
0.18--0.20 & 0.2552 & 0.2531 & 0.2462 & 0.2452 & 0.2557 \\
0.20--0.22 & 0.2861 & 0.2840 & 0.2757 & 0.2743 & 0.2841 \\
0.22--0.24 & 0.3168 & 0.3147 & 0.3047 & 0.3029 & 0.3134 \\
0.24--0.26 & 0.3510 & 0.3490 & 0.3345 & 0.3330 & 0.3453 \\
0.26--0.28 & 0.3863 & 0.3845 & 0.3688 & 0.3673 & 0.3807 \\ 
0.28--0.30 & 0.4377 & 0.4360 & 0.4079 & 0.4071 & 0.4192 \\
0.30--0.32 & 0.4675 & 0.4665 & 0.4370 & 0.4316 & 0.4562 \\
0.32--0.34 & 0.5274 & 0.5257 & 0.4861 & 0.4784 & 0.4955 \\
0.34--0.36 & 0.5430 & 0.5425 & 0.5111 & 0.4961 & 0.5345 \\
\hline
\end{tabular}
\end{table}

\label{lastpage}

\end{document}